\def\@testdef #1#2#3{%
\def\reserved@a{#3}\expandafter \ifx \csname #1@#2\endcsname
\reserved@a  \else
\typeout{^^Jlabel #2 changed:^^J%
\meaning\reserved@a^^J%
\expandafter\meaning\csname #1@#2\endcsname^^J}%
\@tempswatrue \fi}

\documentclass[%
 reprint,
superscriptaddress,
 amsmath,amssymb,
 aps,
]{revtex4-1}

\usepackage{graphicx}
\usepackage{dcolumn}
\usepackage{appendix}
\usepackage{hyperref}
\usepackage[dvipsnames]{xcolor}
\usepackage{comment}
\usepackage{bm}
\usepackage{xr}

\newcommand{\plaind}{\mathrm{d}}
\newcommand{\vecr}{\mathbf{r}}
\newcommand{\vecx}{\mathbf{x}}
\newcommand{\vecJ}{\mathbf{J}}

\newcommand{\fullstop}{\; .}
\newcommand{\comma}{\; ,}

\usepackage{dsfont}

\newcommand{\canetset}[1]{{\mathchoice {\hbox{$\sf\textstyle #1\kern-0.4em #1$}}
{\hbox{$\sf\textstyle #1\kern-0.4em #1$}}
{\hbox{$\sf\scriptstyle #1\kern-0.3em #1$}}
{\hbox{$\sf\scriptscriptstyle #1\kern-0.2em #1$}}}}

\def\nbZ{{\mathchoice {\hbox{$\sf\textstyle Z\kern-0.4em Z$}}
{\hbox{$\sf\textstyle Z\kern-0.4em Z$}}
{\hbox{$\sf\scriptstyle Z\kern-0.3em Z$}}
{\hbox{$\sf\scriptscriptstyle Z\kern-0.2em Z$}}}}

\usepackage{bm}

\newcommand{\Exp}[1]{\operatorname{exp}\left(#1\right)}

\newcommand{\Eref}[1]{Eq.~(\ref{eq:#1})}

\newcommand{\sref}[1]{Sec.~\ref{sec:#1}}
\newcommand{\Sref}[1]{Section~\ref{sec:#1}}

\newlength \standardfigwidth
\newcounter{exercise}
{\addtocounter{exercise}{1}\begin{center}\begin{minipage}{0.8\linewidth}\textbf{Exercise
\arabic{exercise}:}\begin{itshape}}
{\end{itshape}\end{minipage}\end{center}}

\makeatletter
\newcommand{\creat}[3][]{\@ifempty{#1}{#2^{\dagger}}{\left(#2^{\dagger}\right)^{#1}}\@ifempty{#3}{}{\!(#3)}}

\newcommand{\creatDoi}[3][]{\@ifempty{#1}{\tilde{#2}}{\left(\tilde{#2}\right)^{#1}}\@ifempty{#3}{}{(#3)}}

\newcommand{\annih}[3][]{#2\@ifempty{#1}{}{^{#1}}\@ifempty{#3}{}{(#3)}}

\makeatother

% ...

\newlength{\bibmarkkeyAleft}

\newlength{\bibmarkkeyBleft}

\newlength{\bibmarkkeyCleft}

\newlength{\bibmarkkeyDleft}

\makeatletter
\newcommand*{\addFileDependency}[1]{
\typeout{(#1)}
\@addtofilelist{#1}
\IfFileExists{#1}{}{\typeout{No file #1.}}
}\makeatother

\newcommand*{\myexternaldocument}[1]{%
\externaldocument{#1}%
\addFileDependency{#1.tex}%
\addFileDependency{#1.aux}%
}
\myexternaldocument{SM}

\begin{document}

\preprint{APS/123-QED}

\title{
Memoryless Chemotaxis with Discrete Cues
}


\author{Jacob Knight}
\email{jwk21@ic.ac.uk}
\affiliation{Department of Mathematics, Imperial College London, South Kensington, London SW7 2BZ, United Kingdom}

\author{Paula García-Galindo}
\affiliation{Department of Chemical Engineering and Biotechnology, University of Cambridge, Philippa Fawcett Drive, Cambridge CB3 0AS, United Kingdom}

\author{Johannes Pausch}
\affiliation{Department of Mathematics, Imperial College London, South Kensington, London SW7 2BZ, United Kingdom}

\author{Gunnar Pruessner}
\affiliation{Department of Mathematics, Imperial College London, South Kensington, London SW7 2BZ, United Kingdom}

\date{\today}

\begin{abstract}
A wide array of biological systems can navigate in shallow gradients of chemoattractant with remarkable precision. Whilst previous approaches model such systems using coarse-grained chemical density profiles, we construct a dynamical model consisting of a chemotactic cell responding to \textit{discrete} cue particles. For a cell without internal memory, we derive an effective velocity with which the cell approaches a point source of cue particles. 
We find that the effective velocity becomes negative beyond some homing radius, which represents an upper bound on the distance within which chemotaxis can be reliably performed. This work lays the foundation for the analytical characterisation of more detailed models of chemotaxis.
\end{abstract}

\maketitle

\section{Introduction}
To sense its environment, a cell is limited to measuring the occupation of the receptors distributed over its surface \cite{berg_physics_1977}. Using this measurement, the cell can detect diffusive particles which allows it to perform chemotaxis: navigation in an environment with cue chemicals which may attract or  repel the cell \cite{keller_model_1971}. Chemotaxis is an important biological function in single cells since it allows the movement towards a more favourable region for survival and growth \cite{hunter_part_2010}. In multicellular organisms it ensures that cells are in the right place at the right time, which is essential for basic processes such as wound healing \cite{hunter_part_2010}. Additionally, unwanted or unregulated chemotaxis can become a contributing factor in diseases such as cancer, asthma or arthritis \cite{hunter_part_2010}. Chemotaxis can also be performed by cell parts such as the growth cone of neurons where it is called axon guidance \cite{baier_axon_1992}.

A cell's ability to sense gradients from the stochastic arrival of ligands on its receptors has an impact on how efficiently it performs chemotaxis. The physical limits of this process were first calculated for concentration sensing \cite{berg_physics_1977} and later for gradient sensing \cite{goodhill_theoretical_1999, endres_accuracy_2008}. How the measurement of a gradient is performed differs between cell types. In eukaryotic cells the measurement is commonly done directly, where a single cell measures the gradient across the diameter of its body \cite{10.1016/j.ceb.2007.11.011}. This ability of spatial gradient sensing by eukaryotic cells has been shown experimentally in different cells such as slime mold Dictyostelium discoideum (Dicty) \cite{mato_signal_1975, haastert_biased_2007}, the yeast Saccharomyces cerevisiae \cite{arkowitz_responding_1999} and immune system cells such as neutrophils and leukocytes \cite{zigmond_ability_1977}. The cells are remarkably capable of sensing small differences in particle flux, operating close to the physical limits. For example,  neutrophils and Dicty cells can chemotax when there is a concentration difference of only 1\% across the cell length \cite{haastert_biased_2007}. Furthermore, neuronal growth cones can  perform axon guidance by detecting concentration differences of a few per cent across the cone \cite{baier_axon_1992}. 

Chemosensing is a ubiquitous yet complex process in single cell biology. The definition of `sensing' and the parameters involved will differ for each biological context \cite{mortimer_growth_2008}. To study limits of positive chemotaxis we analyse a rudimentary chemotactic strategy. Our model cell has perfect single-molecule sensitivity (each ligand is recorded), moves at constant speed, and instantaneously reorients in the direction of the most recently absorbed ligand. The latter property is reminiscent of the greedy algorithm in which the locally optimal solution is followed \cite{greedy_algo_def}.  As the greedy algorithm results in the best possible response towards finding the source in the absence of information processing and memory, we consider our model as an optimal chemotaxis strategy for a memoryless cell exposed to discrete cue particles.

The paper is organised in the following way. We first define a model cell inspired by Endres and Wingreen's perfectly absorbing sphere \cite{endres_accuracy_2008} that moves using the greedy algorithm, as well as the surrounding environment of diffusive particles. Next, we analytically derive an effective cell velocity which represents an ensemble average of the greedy cell dynamics. We find a threshold distance beyond which an ensemble of cells will move away from the source on average, which we term the \textit{homing radius}. We validate our model by comparing the dynamics of numerical simulations to analytic predictions in different environment regimes. Finally, we discuss the biological significance of our model as well as its limitations, and suggest directions for future research.

\begin{figure}[ht]
    \centering
    \includegraphics[width=0.5\textwidth]{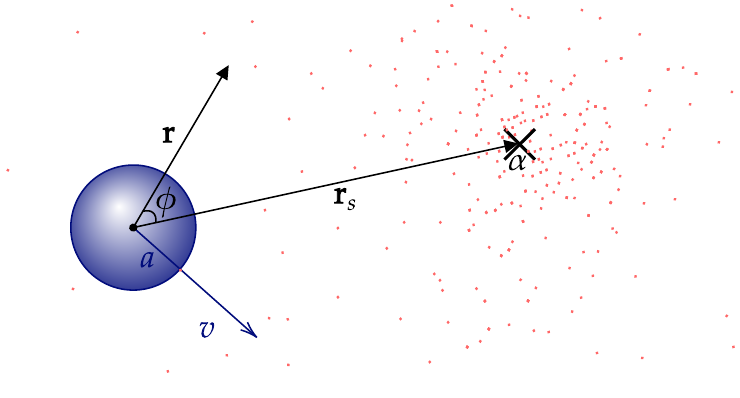}
    \caption{Schematic of the model. The center of the cell (dark blue) is the origin of the coordinate system. Point-like cue particles (red dots) are released with Poissonian rate $\alpha$ from a point source located at $\mathbf{r}_s$ and diffuse with diffusion constant $D$. The cell of radius $a$ absorbs cue molecules on contact and moves with constant speed $v$ in the direction of incidence of the last cue molecule, as illustrated. The vector $\mathbf{r}$ denotes a position where the density of cue particles is evaluated. The angle $\phi \in [0,\pi]$ denotes the polar angle between $\mathbf{r}$ and $\mathbf{r}_s$.}
    \label{fig:schematic}
\end{figure}

\section{Methods}\label{sec:greedy_algorithm}
In this section we introduce and analytically characterise our model of a chemotactic cell in three dimensions, whose motion is governed by the greedy algorithm.
\subsection{Model}\label{sec:model_and_algo}

Consider a spherical cell of radius $a$ in the presence of a point source which releases point-like cue particles with Poissonian rate $\alpha>0$. 
Cue particles represent molecules of a  generic chemoattractant and diffuse with diffusivity $D$.
The cell moves with constant speed $v$ and does not experience rotational or translational diffusion.
The cell can detect cue particles when they collide with its surface. 
Cue particles are absorbed upon contact with the cell surface, as in the \textit{perfectly absorbing sphere} model considered in \cite{endres_accuracy_2008}. 
The cell has no memory, \textit{i.e.} no capacity to store information.
The process terminates successfully if the cell reaches the source.
This setup is illustrated in Fig.~\ref{fig:schematic}.

The motion of the cell in time is determined by the way it responds to incident cue molecules. Given that the cell has no memory, any action it takes must be local in time. In optimisation problems, a greedy algorithm is one which always takes the best immediate, or local, solution \cite{greedy_algo_def}. The optimal generic strategy for the memoryless cell must therefore be to adopt a greedy algorithm which maximises the probability of reaching the source of diffusive particles. Without the benefit of memory, the optimal instantaneous response to receiving a cue particle is to move in the direction of incidence of the most recently absorbed cue particle.

\begin{figure}
    \centering
    \includegraphics[width=0.5\textwidth]{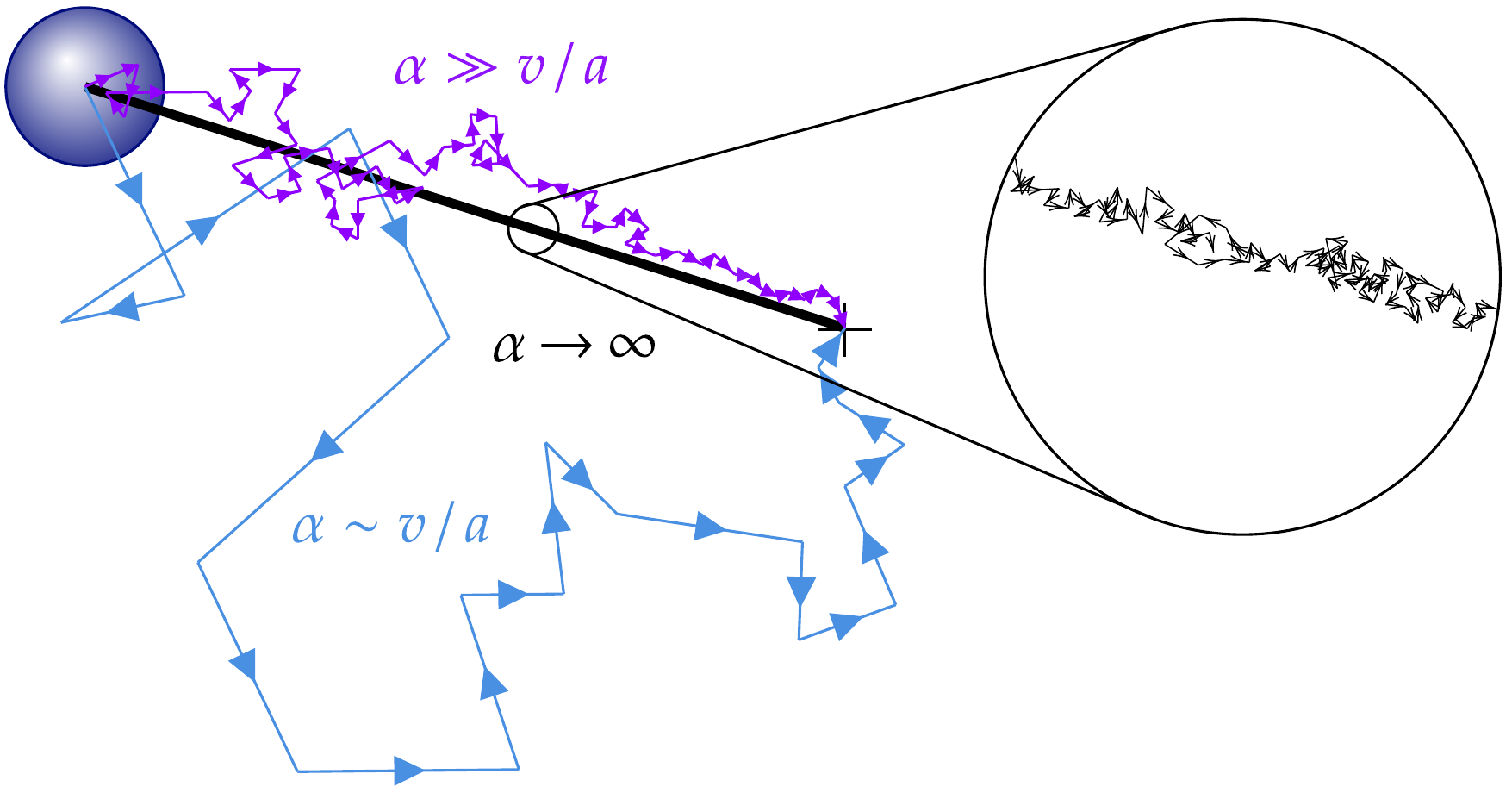}
    \caption{Schematic of trajectories at various cue particle release rates. Runs between collisions with cue particles are longer at low release rate $\alpha$ since the density of cue particles is lower. As the release rate tends to infinity ($\alpha\rightarrow\infty$), runs become infinitely short and the cell trajectory tends towards a straight line. Labels compare the magnitude of the release rate $\alpha$ to the time taken for the cell to move by a single radius, $a/v$. The ratio of these timescales $\varepsilon = \alpha a/v$ characterises the cell trajectory, as derived in section \Sref{finite_release_rate}. }
    \label{fig:traj_schematic}
\end{figure}

The motion of the cell is deterministic in the limit $\alpha \to \infty$ since the probability distribution of cue arrivals on the cell surface is exhaustively sampled and the time between collisions with cues vanishes. 
Conversely, for finite $\alpha>0$, the cell will undertake a random walk whose properties are determined by the release rate $\alpha$ of the cue particles, and the velocity $v$ and radius $a$ of the cell. 
These trajectories are illustrated schematically in Fig.~\ref{fig:traj_schematic}.
We subsequently construct an effective velocity $v_{\rm eff}$ which describes the speed with which the cells approach the source.
 

The cue particle density $\rho(\mathbf{r})$ at position $\mathbf{r}$  due to a source with release rate $\alpha$ at $\mathbf{r}_s$ is given as the solution to Poisson's equation
\begin{align}\label{eq:poisson_only}
    -D\nabla^2\rho(\mathbf{r}) &= \alpha \delta (\mathbf{r}_s - \mathbf{r})\;,
\end{align}
under the boundary conditions:
\begin{subequations}
\label{eq:BCs_only}
    \begin{align}
                \rho(|\mathbf{r}|=a) &= 0,\label{eq: BC_1}\\
\lim_{|\mathbf{r}| \to \infty}\rho(\mathbf{r}) &= 0.\label{eq: BC_2}
    \end{align}
\end{subequations}
Boundary condition~\eqref{eq: BC_1} enforces the complete absorption of cue particles on the surface of the cell whereas boundary condition~\eqref{eq: BC_2} allows cue particles to increasingly dilute away from the source. 

The solution to Eq.~\eqref{eq:poisson_only} which satisfies the boundary conditions~\eqref{eq:BCs_only} is identical (in the region $|\mathbf{r}| > a$) to the solution of:
\begin{subequations}
\label{eq:3d_image}
    \begin{align}
                -D\nabla^2\rho(\mathbf{r}) &= \alpha \delta (\mathbf{r}_s - \mathbf{r}) + \alpha' \delta (\mathbf{r}'_s - \mathbf{r}),\\ \lim_{|\mathbf{r}| \to \infty}\rho(\mathbf{r}) &= 0.\label{eq:BC_3D_one}
    \end{align}
\end{subequations}
The boundary condition (\ref{eq: BC_1}) on the diffusion equation (\ref{eq:poisson_only}) has been implemented by introducing a so-called ``image charge," whose strength $\alpha' = - \alpha \cdot a / r_s $ and position $\mathbf{r}'_s = a^2/r_s^2 \cdot \mathbf{r}_s$ are chosen such that (\ref{eq: BC_1}) is satisfied \cite{barton_elements_1989}. The solution is
\begin{equation}\label{eq:3d_soln}
\begin{split}
    \rho(\mathbf{r})=\,\frac{\alpha}{4\pi D}\Bigg[&\frac{1}{\left(r^2+r_s^2-2\mathbf{r}\cdot\mathbf{r}_s\right)^{\frac{1}{2}}} \\
    &- \frac{1}{\left(r^2r_s^2/a^2+a^2-2\mathbf{r}\cdot\mathbf{r}_s\right)^{\frac{1}{2}}}\Bigg].
\end{split}
\end{equation}
The flux into the cell is proportional to the gradient of the density evaluated at the cell surface and depends only on the polar angle $\phi \in [0,\pi]$ of the considered point $\mathbf{r}$ on the cell surface with respect to the direction of the source $\hat{\mathbf{r}}_s$:
\begin{equation}
\begin{split}\label{eq:flux_3d}
    J(\phi) &= - D \partial_r\rho(\mathbf{r})|_{r=a} \\
    &=\,\frac{\alpha}{4 \pi a}\cdot\frac{r_s^2-a^2}{\left(r_s^2+a^2-2ar_s\cos\phi\right)^{\frac{3}{2}}},
\end{split}
\end{equation}
valid for $r_s>a$ such that the source resides outside the cell, and independent of the diffusivity $D$ as discussed further in \sref{flux_D_independence}.
The rate of cue molecule arrivals through the area of the cell surface between angles in the range $[\phi, \phi + \plaind \phi]$ with respect to the source is given by $J(\phi) \plaind A(\phi)$, where $\plaind A(\phi)=2\pi a^2 \sin(\phi) \plaind \phi$ is the area of the cell surface with a polar angle in the range $[\phi, \phi + \plaind \phi]$.
Once appropriately normalised by the total rate of arrivals of cues at the cell surface, $\int \plaind \phi J(\phi) A(\phi) = \alpha a/r_s$, this gives a probability density for the angle of arrival of cue molecules on the cell surface,
\begin{equation}\label{eq:distn_3d}
    p(\phi) = \frac{r_s}{2} \frac{(r_s^2-a^2)\sin(\phi)}{\left(r_s^2+a^2-2ar_s\cos\phi\right)^{\frac{3}{2}}},
\end{equation}
with $r_s>a$. 

\subsection{Effective velocity at infinite release rate}

In the limit $\alpha \to \infty$, the distribution of cue particles tends towards a smoothly-varying field. Since the total rate of arrivals into the cell $\alpha a / r_s$ also goes to infinity, the time between cue particle arrivals vanishes and the zig-zag motion of the cell becomes a straight line as illustrated in Fig. \ref{fig:traj_schematic}. Its effective velocity towards the source, $v_{\rm eff, \infty}(r_s) = - \plaind r_s / \plaind t$ is the projection of the cell motion in the direction of the source,
\begin{equation}
\begin{split}\label{eq:v_eff_3d}
    v_{\rm eff, \infty}(r_s) &= \mathbb{E}[v\cos\phi]=v\int_{0}^{\pi}\cos(\phi)p(\phi)\plaind\phi, \\
    &= \frac{a v}{r_s}.
    \end{split}
\end{equation}
Since this describes the deterministic motion of the cell, $v_{\rm eff, \infty}(r_s)$ can be integrated to obtain the long-time cell trajectory 
\begin{equation}\label{eq:trajectory_3d}
    \begin{split}
	    r_s(t) = \sqrt{r_0^2 - 2 a v t}\;,
    \end{split}
\end{equation}
where $r_0$ $(\ge a)$ is the initial distance between the source and the centre of the cell.
\\

\subsection{Effective velocity at finite release rate}\label{sec:finite_release_rate}

When cue particles are released at a finite rate $\alpha$, there is a random, finite time between cue arrivals on the cell surface. As such, the executes a random walk, as illustrated in Fig.~\ref{fig:traj_schematic}. 
In the following, we develop the notion of the resulting effective $v_{\rm eff, \alpha}(r_s)$. We consider a cell located at distance $r_s$ from the source, which receives a cue particle at a polar angle $\phi$ relative to the source. 
The random variable describing the time before the next collision with a cue particle is denoted $\Delta t$. Each of these journeys along a straight line are referred to as a ``run'' in the following. The change in radial distance of the cell to the source during a single run is denoted by $\Delta r_s$:
\begin{equation}\label{eq:delta_r_def}
    \Delta r_s(r_s,\phi, \Delta t) = \sqrt{r_s^2 + v^2 \Delta t^2 - 2 r_s v \Delta t\cos(\phi)} - r_s\;.
\end{equation}
The effective velocity for a finite release rate of cue particles can be related to the chemotactic index \cite{endres_accuracy_2008, vanHaastert2007} of a single run. The chemotactic index is defined as the distance moved in the direction of the source divided by the total distance moved by the cell \cite{endres_accuracy_2008},
\begin{equation}\label{eq:CI_def}
    \mathrm{CI} = \frac{\overline{\Delta r_s}}{v \cdot \overline{\Delta t}} \; ,
\end{equation}
where the expectation values $\overline{\bullet}$ are taken with respect to the angular distribution of cue arrivals on the cell surface and the probability distribution of run durations,
\begin{equation}\label{eq:delta_r_bar_def_2}
    \overline{\Delta r_s}(r_s) = \int_0^\pi \mathrm{d}\phi \int_0^\infty \mathrm{d}\Delta t \, \Delta r_s(r_s,\phi, \Delta t) \, p(\phi, \Delta t) \;,
\end{equation}
where $p(\phi, \Delta t)$ denotes the joint probability of a cue at polar angle $\phi$ resulting in a run of duration $\Delta t$. At finite release rate of cue particles (and thus finite incidence rate on the cell surface), the velocity of the cell towards the source is a random number in the range $[-v, v]$. We define its effective value $v_{\rm eff, \alpha}(r_s)$ via the chemotactic index $v_{\rm eff, \alpha}(r_s) = - v \; \mathrm{CI} = - \overline{\Delta r} (r_s) / \overline{\Delta t}(r_s)$. 

To evaluate $\overline{\Delta r} (r_s)$ and $\overline{\Delta t} (r_s)$, we express the joint probability density $p(\phi, \Delta t) = p(\phi)p(\Delta t|\phi)$ as the product of the angular distribution (\ref{eq:distn_3d}) and the distribution of the run duration $\Delta t$ conditioned on a collision at an angle $\phi$.
Since the distribution of cue particles is inhomogeneous, cue-cell collisions are governed by a Poisson process with rate dependent on the position of the cell relative to the source. The rate of cue collisions with the cell $\lambda\big(r_s\big)$ is equal to the total flux of cue particles into the cell, $\lambda\big(r_s\big) = \int \plaind \phi J(\phi) A(\phi) = \alpha a/r_s$. For a cell which experiences a collision occurring at a distance $r_s$ and polar angle $\phi$ relative to the source and continues to move for a further time $t$, the total rate of cue arrivals as a function of time becomes
\begin{equation}\label{eq:lambdadef2}
    \lambda\big(t\big) = \frac{\alpha a}{\sqrt{r_s^2 + v^2 t^2 - 2 r_s v t\cos(\phi)}}\;.
\end{equation}
The probability density of $\Delta t$ conditioned on cue arrival at an angle $\phi$ is hence given by 
\begin{equation}\label{eq:deltat_cond_prob}
    p(\Delta t|\phi) = \lambda(\Delta t) \Exp{ - \int_0^{\Delta t} \mathrm{d} t \lambda (t)}\;,
\end{equation}
which is calculated explicitly in App. \ref{SM:veff_derivation}. 
With the probability densities known, the mean change in radial coordinate $\overline{\Delta r_s}$ and the mean run duration $\overline{\Delta t}$ can be expressed in terms of a dimensionless parameter $\varepsilon = \frac{\alpha a}{v}$ (App. \ref{SM:veff_derivation}),
\begin{subequations}\label{eq:step_results}
   \begin{align}
	   \overline{\Delta r_s}(r_s) &= \frac{a \varepsilon - r_s}{1 - \varepsilon^2}  \;, \\
			   \overline{\Delta t}(r_s) &= \frac{1}{v}\; \frac{\varepsilon r_s - a}{1 - \varepsilon^2} \;.
   \end{align}
\end{subequations}
These results are valid only for $\varepsilon > 1$ as both $\overline{\Delta r}(r_s)$ and $\overline{\Delta t}(r_s)$ diverge for $\varepsilon \leq 1$. 
This condition has a physical significance which is discussed in Sec. \ref{sec:discussion}.
From Eq.~\eqref{eq:CI_def}, the chemotactic index of the cell is
   \begin{align}
	  \mathrm{CI}(r_s) 
			    &= \frac{a \varepsilon - r_s}{r_s \varepsilon - a} \;,\end{align}
and the effective velocity is therefore
   \begin{align}\label{eq:veff_result}
	   v_{\rm eff, \alpha}(r_s) 
			    &= v \  \frac{a \varepsilon - r_s}{r_s \varepsilon - a} \;. \end{align}
The signs of $\overline{\Delta r}(r_s)$ and hence $\mathrm{CI}(r_s)$ and $v_{\rm eff, \alpha}(r_s)$ change at a radius which we term the \textit{homing radius}, 
\begin{equation}\label{eq:homing_rad_def}
    r_h = a \varepsilon = \frac{\alpha a^2}{v}.
\end{equation}
The mechanism by which this occurs is discussed in Sec. \ref{sec:discussion}.

\begin{figure}[t]
    \centering
    \includegraphics[width=\linewidth]{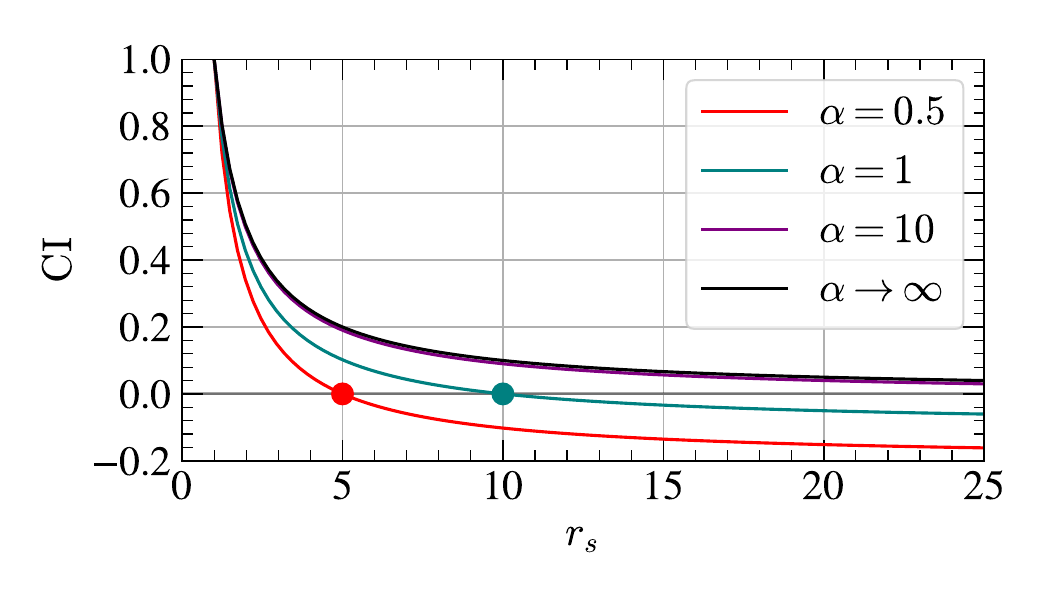}
    \caption{Plot of the chemotactic index $\mathrm{CI}(r_s)$ for various values of $\alpha$, measured in units of $s^{-1}$. For finite values of $\alpha$, there is a homing radius $r_h$ defined in Eq.~\eqref{eq:homing_rad_def} beyond which the chemotactic index, and hence the effective velocity, of the cell is negative, corresponding to a net migration of cells away from the source. The homing radii at $\alpha=0.5,1$ are marked on the plot with solid points. The chemotactic index for finite $\alpha$ defined in Eq.~\eqref{eq:CI_def} converges to that of infinite $\alpha$ as denoted in Eq.~\eqref{eq:v_eff_3d} (noting that $v_{\rm eff}(r_s) = - v \:\mathrm{CI}(r_s)$). For each curve the cell radius and velocity are $a=1$ and $v=0.1$ respectively.}
    \label{fig:v_eff_vs_rs}
\end{figure}

Rather than the deterministic speed $v_{\rm eff, \infty}(r_s)$ of a cell, $v_{\rm eff, \alpha}(r_s)$ instead describes the rate of approach towards the source of the mean position of an ensemble of independent cells (each with an independent set of cue particles) over a single run. 
After one run, cells will be spread out in space with some non-linear distribution. Since Eq.~\eqref{eq:veff_result} applies only to an ensemble at a single point in space and is non-linear in $r_s$, the subsequent motion of cells cannot be described by integrating Eq.~\eqref{eq:veff_result} as it was in Eq.~\eqref{eq:trajectory_3d}. Nonetheless, Eq.~\ref{eq:v_eff_3d} is recovered in the limit $\alpha \to \infty$
\begin{equation}
    \lim_{\alpha \to \infty}  v_{\rm eff, \alpha}(r_s) = \frac{\alpha a}{v} = v_{\rm eff, \infty}(r_s) \;,
\end{equation}
as illustrated in Fig.~\ref{fig:v_eff_vs_rs}. 

\subsection{Comparison with simulation}

We numerically simulated the system described in Sec. \ref{sec:model_and_algo} , with the addition of an outer boundary at a large distance $R_c \gg a,r_s, r_0$ as a cutoff at which a cell was deemed irretrievably lost on its way to the source. This outer boundary also absorbed cue particles, which was necessary for the simulation to reach a steady state. The deviation from the density distribution \Eref{3d_soln} that this caused was found to be negligibly small. 

The system was initialised without cue particles, with the absorbing cell placed a distance $r_s = r_0$ from the source. Cue particles were subsequently released from the source at rate $\alpha$ and diffused with diffusion constant $D$. The system was allowed sufficient time to reach a steady state with the cell remaining stationary, \textit{i.e.} not responding to cues. After this time and once the first cue arrived, the clock was reset to $t=0$ and the cell began to move according to the greedy algorithm (Sec. \ref{sec:model_and_algo}), and its position recorded as a function of time. Each trial terminated when the cell reached the source or the outer boundary. Upon termination of the trial, the cell was reset to its initial position, without changing the positions of cue particles. Sufficient time was allowed for the system to return to its original steady state distribution, with the cell stationary. This was repeated 1,000 times in each simulation. 

\begin{figure*}[ht]
    \centering
    \includegraphics[width=\linewidth]{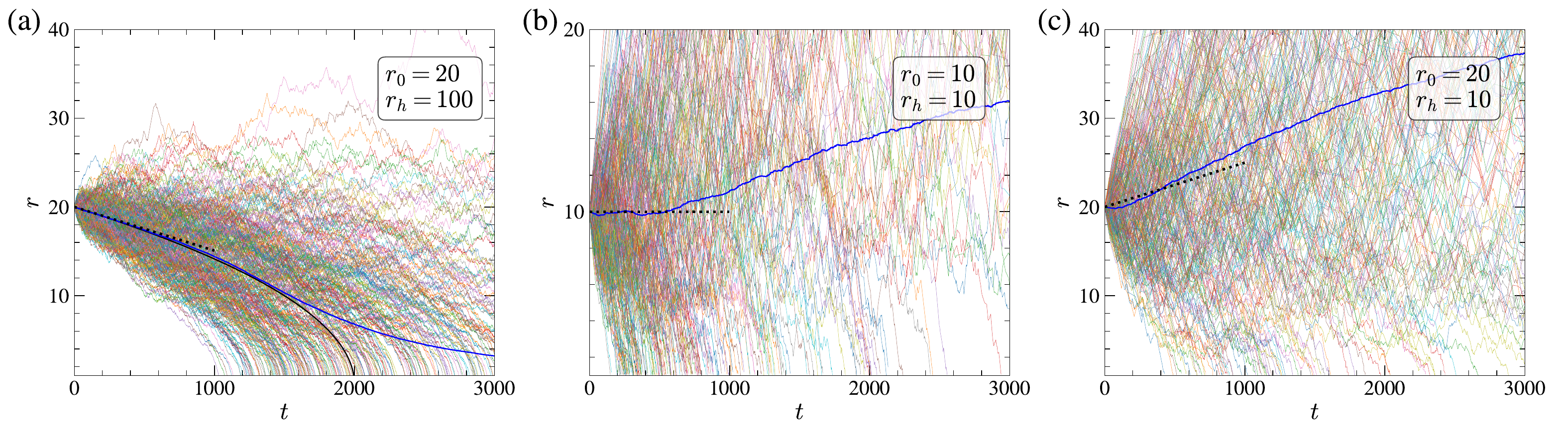}
    \caption{Sample trajectories of ensembles of cells initialised at $r_0<r_h$, $r_0=r_h$ and $r_0>r_h$ respectively. Multicoloured lines represent the trajectories of individual cells, while the blue solid line denotes the mean distance of the cells from the source averaged over 1000 trajectories. The black solid line in panel (a) shows the trajectory described by Eq.~\eqref{eq:trajectory_3d} for infinite release rate and the black dotted lines in all panels show the predicted instantaneous velocity from Eq.~\eqref{eq:veff_result}. The initial motion of the center of mass of the cells is described well by Eq.~\eqref{eq:trajectory_3d}. Qualitatively, the observed long-time positive deviation from theoretical predictions occurs because some proportion of the initial ensemble runs away from the source towards large $r$. This part of the ensemble dominates the mean particle position in the long-time regime since they reach increasingly large $r$.  
    For all runs, $v=0.1$ and $a=1$. The other parameters are: (a) $r_0 = 20$, $\alpha= 10$, (b) $r_0 = 10$, $\alpha= 1$, (c) $r_0 = 20$, $\alpha= 1$. }
    \label{fig:traj_tryptich}
\end{figure*}

Sample trajectories $\{r_s(t)\}$ from this simulation are shown in Fig.~\ref{fig:traj_tryptich}. The simulations confirm that the effective velocity defined in Eq.~\eqref{eq:veff_result} successfully predicts the initial velocity of the ensemble-averaged position of the cells. It also demonstrates that Eq.~\eqref{eq:trajectory_3d} provides a good approximation of the long-term trajectory of the center of mass of an ensemble of particles with large but finite release rate. 

Deviations of the numerics from the predicted behaviour have two causes. Firstly, we consider the motion of an ensemble of cells all starting at a single point, not accounting for their subsequent spread in space according to the independent collision events for each cell. Secondly, our theory does not account for correlations in the location of the cue particles due to the ``wake" left behind by the cell. These two points are further discussed in the following section.

\section{Discussion}\label{sec:discussion}

\subsection{Chemotaxis in arbitrarily shallow gradients }\label{sec:flux_D_independence}

One of the remarkable features of eukaryotic cell behaviour is their ability to sense gradients as shallow as 1--5\% across their body length \cite{mato_signal_1975}. Understanding the mechanism by which this is achieved is a long-standing open question in the study of chemotaxis. We make the observation in our model that the chemical gradient in which the cell navigates can be made arbitrarily shallow without affecting chemotactic performance. 

From \Eref{3d_soln}, the local chemical gradient at a cell's surface is proportional to the chemoattractant release rate divided by the diffusivity, $\mathbf{\nabla}_\vecr \rho(\vecr) \rvert_{\vecr = \vecx} \propto \alpha/D$. However,  \Eref{flux_3d} shows that the flux into the cell is proportional to the release rate but independent of the diffusivity, $\vecJ(\vecx) \propto \alpha$. As such, the chemical gradient in the vicinity of a cell can be made arbitrarily small by increasing the diffusivity $D$ without affecting the flux incident on the cell and hence its ability to perform chemotaxis. Intuitively, increasing $D$ will spread out the chemoattractant profile, thereby decreasing the local density around the cell. This is compensated for by the higher mean-square displacement of nearby cue particles which enter the cell more frequently.

Existing literature posits that cells can sense gradients either by measuring the chemical \textit{flux} into their surface or the chemical density inside their volume \cite{endres_accuracy_2008}. 
We argue that the diffusion-independence of particle flux into the cell surface provides a physical explanation for the ability of flux-sensing cells to navigate in extremely shallow gradients. This is not only true in our model of a spherical cell; we demonstrate in App.~\ref{SM:flux_diffusivity} that molecular flux into a cell is independent of diffusivity \textit{for cells of any shape} and for \textit{any arrangement of sources of chemoattractant}, providing a convincing explanation for the ability of cells to navigate in shallow chemical gradients.

\subsection{Sign change in effective velocity}

\begin{figure}[ht]
    \centering
    \includegraphics[width=0.85\linewidth]{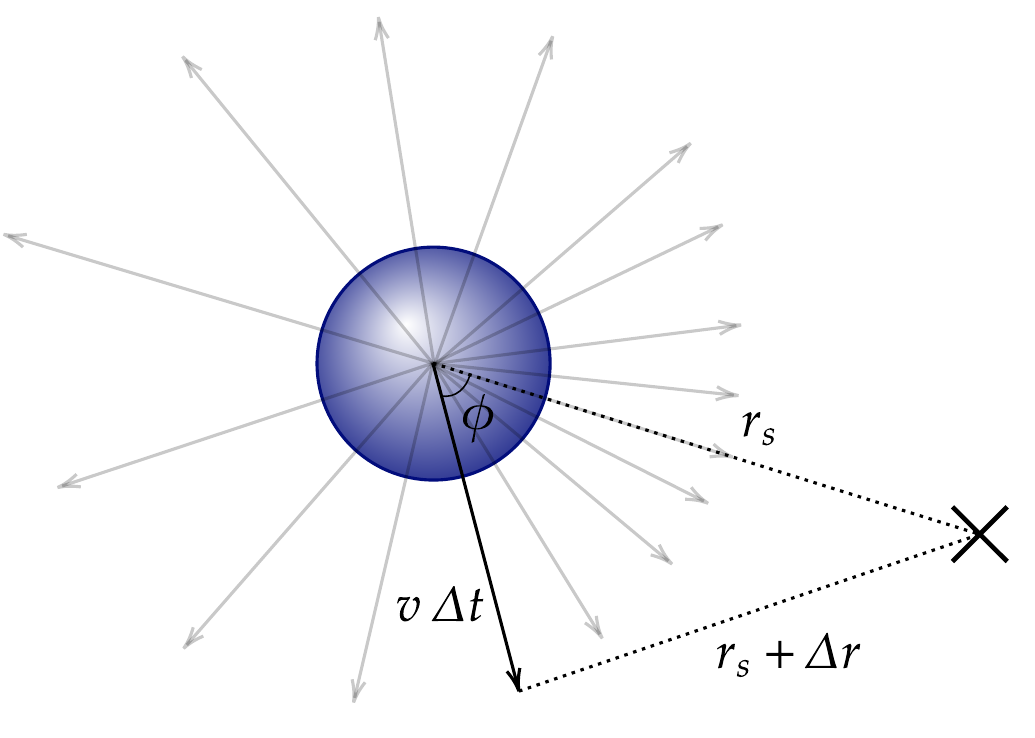}
    \caption{Schematic illustration of runs at finite $\alpha$. Although the angular distribution of runs is biased towards the source, runs away from the source tend to be longer since the density of cue particles is lower further from the source. Beyond the homing radius $r_h$, this leads to a negative chemotactic index and hence a net migration of cells away from the source.}
    \label{fig:boiled_egg}
\end{figure}

When $\alpha$ is finite, there is a homing radius defined in Eq.~\eqref{eq:homing_rad_def} beyond which the chemotactic index and hence the effective velocity are negative. Two competing factors affect the effective velocity of the cell: the bias of the angular distribution Eq.~\eqref{eq:distn_3d} towards the source, and the fact that runs away from the source tend to be longer. 
The bias of the angular distribution is visibile by rescaling $p(\phi)$ in Eq.~\eqref{eq:distn_3d} to be the density per solid angle, $p(\phi)/\sin(\phi)$, which has its maximum at $\phi=0$ and its minimum at $\phi=\pi$. The competing effect is that the position-dependent cue collision rate $\lambda(r)\propto 1/r$ decreases with distance from the source. 
As such, runs away from the source persist for longer. 
The cell is less likely to move away from the source than towards it, but if it does, it takes longer for a cue particle to collide with it and change its direction. 
This effect is illustrated schematically in Fig.~\ref{fig:boiled_egg}. 

At the homing radius, these effects cancel out, leading to no net movement of the cell on average. Of an ensemble of cells initialised at $r_s = r_h$ and allowed to complete their trajectories, more than half would move in the direction of the source over their first run. This is because the bias in angular distribution of cue arrivals means that more than half of the cells would take an initial step towards the source, into a region where $v_{\rm eff, \alpha} > 0$. To calculate the proportion of cells which eventually succeed in arriving at the source is challenging because it involves correlated sequences of runs. An outlined derivation to be explored in future work is given in App. \ref{SM:radial_transition_prob}. 

The homing radius is derived solely from microscopic system parameters. 
This stands in contrast to previous approaches, which require the introduction of an external cutoff on the signal-to-noise ratio of the arrivals on the cell surface \cite{hein_physical_2016}.
Our model also makes a concrete prediction about the effective velocity of the cell.
The homing radius further provides a useful benchmark against which experimental observations of chemotaxis can be compared. 

\subsection{Range of validity of results}\label{sec:validity_of_results}

The condition $\varepsilon>1$ which is required to obtain Eqs.~(\ref{eq:step_results}) to (\ref{eq:veff_result}) has a number of physical interpretations. Multiplying both sides of the inequality by $a$ yields $a \varepsilon = r_h > a$, i.e. the cell radius must be strictly less than the homing radius of the system. If this condition was broken,  cells could be initialised at $r_s = a \geq r_h$, and would instantly be in contact with the source despite being outside the homing radius, which should imply that they move away from the source on average. Even if the cell touches the source, it would move away on average since $a>r_h$. 

The parameter $\varepsilon$ can be expressed as the ratio of two timescales, $\varepsilon = \alpha a / v = \tau_b / \tau_r$, where $\tau_b = a/v$ is the time taken for the cell to more a distance equal to its radius and $\tau_r = 1/\alpha$ is the mean time between the release of cue particles. As such, requiring $\varepsilon>1$ corresponds to demanding a sufficiently high cue release rate that the cell does not travel more than one radius in the mean time between releases.

Intuition can also be gained from the fact that the probability distribution of run durations is governed by a power law: as $\Delta t \to \infty$, $p(\Delta t) \propto \Delta t ^{-1-\varepsilon}$. It is evident that the integral $\overline{\Delta t} = \int _0^\infty \mathrm{d}\Delta t \Delta t p(\Delta t)$ only converges for $\varepsilon > 1$. Similar reasoning applies to $\overline{\Delta r}$. In qualitative terms, if the cell runs at an angle $\phi = \pi$, and $\varepsilon \leq 1$, the expected time to encounter a cue particle diverges. As such, the expected step duration and length diverge. 

\subsection{Model assumptions}
Several assumptions made in this model are stated in section \ref{sec:model_and_algo}. 
One assumption is that the cell is not affected by translational or rotational diffusion. 
We expect that cell diffusion can be superimposed and would decrease the chemotactic index, corresponding to impaired cell performance. This could be implemented by replacing the denominator of \Eref{lambdadef2} with the expected radial coordinate of an appropriately initialised active Brownian particle. Although mathematically feasible, this lies beyond the scope of the current work. Furthermore, values of initial distances, velocities and diffusion constants realistic to eukaryotic cells produce Péclet numbers in the thousands, meaning that corrections are expected to be small.

The model also assumes that the cell reorientation occurs instantaneously. 
In biological systems, rearrangement of cellular motility apparatus takes a non-negligible amount of time \cite{greenfield_self-organization_2009}. We argue that neglecting cell reorientation time is justified in each of two limits in our model. In the limit of sparse cue particles and hence few tumbles, the time taken for reorientation will be small compared to the time spent in motion. In the limit of dense cue particles ($\alpha \to \infty$), the flux into the cell becomes infinite and trajectory of the cell becomes a straight line (see Fig.~\ref{fig:traj_schematic}). The cell would therefore remain stationary if finite time was required for the cell to reorientate. In this limit, the cell moves according to an effective average over arrivals on its surface (despite performing no computation itself), and can be considered to be self-propelling in a single deterministic direction with velocity given by \Eref{v_eff_3d}. Corrections between these limits could be introduced in extensions to the model.

The distribution of cue particles was assumed to be quasistatic, such that the density profile of cue particles in the system is always given by the solution to Eq.~\eqref{eq:3d_image}. This is valid when the timescale for a cue particle to diffuse some distance $x$ is much larger than the time for the cell to self-propel the same distance, equivalent to the condition $D/(a v) \gg 1$. When this condition is not satisfied, the cell leaves a low-density region, or ``shadow," in its wake \cite{pausch_is_2019}. Capturing these correlations is very difficult but neglecting them is justified since cue particles are far smaller than the cell, with diffusivity $D$ orders of magnitude less than that of the cell (and greater than the product $av$, as discussed above). Simulation parameters to generate Fig.~\ref{fig:traj_tryptich} were chosen such that $D/av = 10$ and it was checked that varying $D$ did not significantly affect the statistics of observed trajectories. 

The present model also assumes that cue particles have an infinite lifetime after release. Recent work has shown that the generation of morphogen gradients in early \textit{Drosophila} embryos and eggs is consistent with the synthesis-diffusion-degradation (SDD) model \cite{bicoid, bicoid2}. This corresponds to \Eref{poisson_only} with an additional term to represent the decay of cue particles. Although this equation cannot be solved analytically with the boundary conditions given in \Eref{BCs_only}, a numerical analysis following the same steps would be tractable and an interesting avenue for further investigation.

A limitation of the present theory was that an effective velocity could only be calculated for an  ensemble of cells initialised at the \textit{same} distance from the source, rather than spread out in space. 
After a short time, an ensemble of cells initially at the same distance from the source will be spread out in space, each moving according to the first arriving cue. This makes the subsequent calculation of an effective velocity of the center of mass of the cells difficult, since the effective velocity only applies to an ensemble at the same distance from the source. As a result, we can only exactly predict the long-time trajectories of cells in the limit of $\alpha \to \infty$, where the trajectories are deterministic. For finite $\alpha$, the effective velocity describes the motion of the mean distance of the cells from the source only for a short time $\sim \Delta t$. 


\subsection{Biological context}
In biology, cells establish a sense of distance and direction through chemical cue gradients \cite{lander_how_2013, parent_cells_1999}.
Most existing models, including widely-used works \cite{keller_model_1971, berg_chemotaxis_1972}, treat the concentration of chemoattractant as a continuously-varying field. This commonly used mean-field approach to estimate concentration and gradient neglects the particle nature of chemoattractants. However, the discreteness of chemicals becomes more relevant at the low density regime, in which it is well-established that eukaryotic cells can adopt a favourable direction by considering only a few binding events of chemical cues \cite{haastert_biased_2007, baier_axon_1992}. Even at high chemottractant densities, eukaryotic cells adopt their direction in the timescale of the first binding events \cite{segall_polarization_1993}. Indeed, a simple model has recently shown how the first bindings can confer enough information to accurately estimate the source
direction at a close enough distance \cite{bernoff_single-cell_2023}. 

We analyse the particle nature of chemoattractants through a dynamical and perfectly sensitive cell that has rudimentary features, meaning no intricate processing or memory. The cell optimally searches the source using local information, an approach to understanding chemotaxis similar in spirit to infotaxis \cite{vergassola_infotaxis_2007}. However, the infotaxis strategy considers the exploitation versus exploration trade-off, while the greedy strategy that the cell uses only takes an exploitation approach as it moves instantaneously after the first hit. As a strategy, the greedy cell could be studied in the context of evolutionary adaptation. One can consider this as a hypothetical rudimentary ancestor of more biologically realistic cell chemotactic behaviours that can adapt better to a noisy environment with the evolutionary trade-off of developing more sophisticated processing or memory.

The model presented in this work describes the spatial gradient sensing of eukaryotic cells more closely than the temporal gradient sensing of bacteria. Such bacteria typically execute run-and-tumble dynamics, during which gradient sensing occurs over an extended period of time. However, the model cell in this paper uses local information to direct its movement instantaneously, which relates to eukaryotic direct spatial gradient measurement, where the cell estimates the gradient across the diameter of its body \cite{gerisch_chemotaxis_1982, eidi_modelling_2017}.

The eukaryotic Dicty cell is a particularly well-studied model organism for chemotaxis. They have a typical velocity of 1-2 body lengths per minute, placing them well within the regime $D/av\gg1$ at room temperature. An earlier investigation by Endres and Wingreen \cite{endres_accuracy_2008} provided strong evidence that Dicty cells sense gradients over their surface and move accordingly. The model in this paper extends their results to a regime where concentration is low enough that cue particles are sparse.

An example of such a system is the chemotaxis of immune cells such as macrophages or neutrophils towards a wound infection site \cite{alberts_innate_2002}. Immune cells are densely packed in tissue so that these can reliably detect infections. There are around $10^6$ immune cells per gram of tissue \cite{sender_total_2023} and 1g of tissue is around 1cm$^3$ \cite{del_monte_does_2009}. Therefore, there is around a 100$\mu$m distance between the cells, corresponding to about 5-10 cell diameters \cite{krombach_cell_1997}. If the velocity of the immune cell is measured in units of cell radii per second \cite{de_filippo_secretive_2020}, then the homing radius is the release rate of chemoattractant particles times the radius of the cell. 
If the release rate of a chemoattractant from a wound infection site could be measured and the velocity of the cell is not too large, the corresponding homing radius may be comparable or larger value than the typical distance between the cells previously calculated.
This example suggests that the minimal immune cell density required to reliably detect a wound could be conditioned by the homing radius of the sources to which the immune cell responds.

\section{Conclusion}
\noindent
Statistical fluctuations due to diffusion set a physical limit for concentration \cite{berg_physics_1977} and concentration gradient sensing \cite{endres_accuracy_2008}, which eukaryotic cells in nature approach remarkably closely \cite{haastert_biased_2007, endres_accuracy_2008}. This work explores physical limits of chemotaxis using a novel approach that captures the discrete nature of chemoattractant cue particles, going beyond previous work on gradient sensing. The chemotactic performance of a simple cell with no internal memory is characterised analytically. In the limit of infinite cue particle release rate, we derive the deterministic effective velocity and trajectory of the cell. When the cue particle release rate is finite, the cell behaviour becomes stochastic and we derive an effective velocity capturing the mean motion of an ensemble of cells. A characteristic ``homing radius" $r_h$ emerges, beyond which the cell's chemotactic index becomes negative (i.e. cells move away from the source on average). As well as acting as a benchmark against which the chemotactic performance of cells can be compared experimentally, this model provides novel insight into how strength of a chemical source limits the ability of cells to navigate towards it. This model could be extended in future to add diffusive dynamics and a finite reorientation time to the cell. Questions of further interest would be the extent to which introducing memory and processing capacity to the cell would increase chemotactic performance, and quantitative comparisons between the predictions of the model and real eukaryotic cells. 

\begin{acknowledgements}
We thank Robert Endres for useful discussions and Talia Rahall for her involvement in the project and her contributions to a different chemotaxis model. We also thank Ankit Ranjan and Xinyue Fan for their useful work with G.P. and J.P. on related aspects of chemotaxis.  J.K.\ acknowledges  support from the Engineering and Physical Sciences Research Council (grant number 2620369). P.G-G. acknowledges support from ``La Caixa" Foundation. J.P. was supported through a UKRI Future Leaders Fellowship (MR/T018429/1 to Philipp Thomas).
\end{acknowledgements}

\bibliography{references.bib}

\begin{thebibliography}{33}%
\makeatletter
\providecommand \@ifxundefined [1]{%
 \@ifx{#1\undefined}
}%
\providecommand \@ifnum [1]{%
 \ifnum #1\expandafter \@firstoftwo
 \else \expandafter \@secondoftwo
 \fi
}%
\providecommand \@ifx [1]{%
 \ifx #1\expandafter \@firstoftwo
 \else \expandafter \@secondoftwo
 \fi
}%
\providecommand \natexlab [1]{#1}%
\providecommand \enquote  [1]{``#1''}%
\providecommand \bibnamefont  [1]{#1}%
\providecommand \bibfnamefont [1]{#1}%
\providecommand \citenamefont [1]{#1}%
\providecommand \href@noop [0]{\@secondoftwo}%
\providecommand \href [0]{\begingroup \@sanitize@url \@href}%
\providecommand \@href[1]{\@@startlink{#1}\@@href}%
\providecommand \@@href[1]{\endgroup#1\@@endlink}%
\providecommand \@sanitize@url [0]{\catcode `\\12\catcode `\$12\catcode `\&12\catcode `\#12\catcode `\^12\catcode `\_12\catcode `\%12\relax}%
\providecommand \@@startlink[1]{}%
\providecommand \@@endlink[0]{}%
\providecommand \url  [0]{\begingroup\@sanitize@url \@url }%
\providecommand \@url [1]{\endgroup\@href {#1}{\urlprefix }}%
\providecommand \urlprefix  [0]{URL }%
\providecommand \Eprint [0]{\href }%
\providecommand \doibase [0]{http://dx.doi.org/}%
\providecommand \selectlanguage [0]{\@gobble}%
\providecommand \bibinfo  [0]{\@secondoftwo}%
\providecommand \bibfield  [0]{\@secondoftwo}%
\providecommand \translation [1]{[#1]}%
\providecommand \BibitemOpen [0]{}%
\providecommand \bibitemStop [0]{}%
\providecommand \bibitemNoStop [0]{.\EOS\space}%
\providecommand \EOS [0]{\spacefactor3000\relax}%
\providecommand \BibitemShut  [1]{\csname bibitem#1\endcsname}%
\let\auto@bib@innerbib\@empty
\bibitem [{\citenamefont {Berg}\ and\ \citenamefont {Purcell}(1977)}]{berg_physics_1977}%
  \BibitemOpen
  \bibfield  {author} {\bibinfo {author} {\bibfnamefont {H.}~\bibnamefont {Berg}}\ and\ \bibinfo {author} {\bibfnamefont {E.}~\bibnamefont {Purcell}},\ }\href@noop {} {\bibfield  {journal} {\bibinfo  {journal} {Biophys. J.}\ }\textbf {\bibinfo {volume} {20}},\ \bibinfo {pages} {193} (\bibinfo {year} {1977})}\BibitemShut {NoStop}%
\bibitem [{\citenamefont {Keller}\ and\ \citenamefont {Segel}(1971)}]{keller_model_1971}%
  \BibitemOpen
  \bibfield  {author} {\bibinfo {author} {\bibfnamefont {E.~F.}\ \bibnamefont {Keller}}\ and\ \bibinfo {author} {\bibfnamefont {L.~A.}\ \bibnamefont {Segel}},\ }\href {\doibase 10.1016/0022-5193(71)90050-6} {\bibfield  {journal} {\bibinfo  {journal} {J. theor. biol.}\ }\textbf {\bibinfo {volume} {30}},\ \bibinfo {pages} {225} (\bibinfo {year} {1971})}\BibitemShut {NoStop}%
\bibitem [{\citenamefont {Hunter}(2010)}]{hunter_part_2010}%
  \BibitemOpen
  \bibfield  {author} {\bibinfo {author} {\bibfnamefont {T.}~\bibnamefont {Hunter}},\ }in\ \href@noop {} {\emph {\bibinfo {booktitle} {Handbook of cell signaling (second edition)}}},\ \bibinfo {editor} {edited by\ \bibinfo {editor} {\bibfnamefont {R.~A.}\ \bibnamefont {Bradshaw}}\ and\ \bibinfo {editor} {\bibfnamefont {E.~A.}\ \bibnamefont {Dennis}}}\ (\bibinfo  {publisher} {Academic Press},\ \bibinfo {address} {San Diego},\ \bibinfo {year} {2010})\ \bibinfo {edition} {second edition}\ ed.,\ pp.\ \bibinfo {pages} {385--389}\BibitemShut {NoStop}%
\bibitem [{\citenamefont {Baier}\ and\ \citenamefont {Bonhoeffer}(1992)}]{baier_axon_1992}%
  \BibitemOpen
  \bibfield  {author} {\bibinfo {author} {\bibfnamefont {H.}~\bibnamefont {Baier}}\ and\ \bibinfo {author} {\bibfnamefont {F.}~\bibnamefont {Bonhoeffer}},\ }\href {\doibase 10.1126/science.1734526} {\bibfield  {journal} {\bibinfo  {journal} {Science}\ }\textbf {\bibinfo {volume} {255}},\ \bibinfo {pages} {472} (\bibinfo {year} {1992})}\BibitemShut {NoStop}%
\bibitem [{\citenamefont {Goodhill}\ and\ \citenamefont {Urbach}(1999)}]{goodhill_theoretical_1999}%
  \BibitemOpen
  \bibfield  {author} {\bibinfo {author} {\bibfnamefont {G.~J.}\ \bibnamefont {Goodhill}}\ and\ \bibinfo {author} {\bibfnamefont {J.~S.}\ \bibnamefont {Urbach}},\ }\href {\doibase 10.1002/(SICI)1097-4695(19991105)41:2<230::AID-NEU6>3.0.CO;2-9} {\bibfield  {journal} {\bibinfo  {journal} {J. Neurobiol.}\ }\textbf {\bibinfo {volume} {41}},\ \bibinfo {pages} {230} (\bibinfo {year} {1999})}\BibitemShut {NoStop}%
\bibitem [{\citenamefont {Endres}\ and\ \citenamefont {Wingreen}(2008)}]{endres_accuracy_2008}%
  \BibitemOpen
  \bibfield  {author} {\bibinfo {author} {\bibfnamefont {R.~G.}\ \bibnamefont {Endres}}\ and\ \bibinfo {author} {\bibfnamefont {N.~S.}\ \bibnamefont {Wingreen}},\ }\href {\doibase 10.1073/pnas.0804688105} {\bibfield  {journal} {\bibinfo  {journal} {Proc. Natl. Acad. Sci.}\ }\textbf {\bibinfo {volume} {105}},\ \bibinfo {pages} {15749} (\bibinfo {year} {2008})}\BibitemShut {NoStop}%
\bibitem [{\citenamefont {Iglesias}\ and\ \citenamefont {Devreotes}(2008)}]{10.1016/j.ceb.2007.11.011}%
  \BibitemOpen
  \bibfield  {author} {\bibinfo {author} {\bibfnamefont {P.~A.}\ \bibnamefont {Iglesias}}\ and\ \bibinfo {author} {\bibfnamefont {P.~N.}\ \bibnamefont {Devreotes}},\ }\href {\doibase 10.1016/j.ceb.2007.11.011} {\bibfield  {journal} {\bibinfo  {journal} {Curr. Opin. Cell Biol.}\ }\textbf {\bibinfo {volume} {20}},\ \bibinfo {pages} {35} (\bibinfo {year} {2008})}\BibitemShut {NoStop}%
\bibitem [{\citenamefont {Mato}\ \emph {et~al.}(1975)\citenamefont {Mato}, \citenamefont {Losada}, \citenamefont {Nanjundiah},\ and\ \citenamefont {Konijn}}]{mato_signal_1975}%
  \BibitemOpen
  \bibfield  {author} {\bibinfo {author} {\bibfnamefont {J.~M.}\ \bibnamefont {Mato}}, \bibinfo {author} {\bibfnamefont {A.}~\bibnamefont {Losada}}, \bibinfo {author} {\bibfnamefont {V.}~\bibnamefont {Nanjundiah}}, \ and\ \bibinfo {author} {\bibfnamefont {T.~M.}\ \bibnamefont {Konijn}},\ }\href {\doibase 10.1073/pnas.72.12.4991} {\bibfield  {journal} {\bibinfo  {journal} {Proc. Natl. Acad. Sci. Academy of Sciences}\ }\textbf {\bibinfo {volume} {72}},\ \bibinfo {pages} {4991} (\bibinfo {year} {1975})}\BibitemShut {NoStop}%
\bibitem [{\citenamefont {Haastert}\ and\ \citenamefont {Postma}(2007)}]{haastert_biased_2007}%
  \BibitemOpen
  \bibfield  {author} {\bibinfo {author} {\bibfnamefont {P.~J.~v.}\ \bibnamefont {Haastert}}\ and\ \bibinfo {author} {\bibfnamefont {M.}~\bibnamefont {Postma}},\ }\href {\doibase 10.1529/biophysj.107.104356} {\bibfield  {journal} {\bibinfo  {journal} {Biophys. J.}\ }\textbf {\bibinfo {volume} {93}},\ \bibinfo {pages} {1787} (\bibinfo {year} {2007})}\BibitemShut {NoStop}%
\bibitem [{\citenamefont {Arkowitz}(1999)}]{arkowitz_responding_1999}%
  \BibitemOpen
  \bibfield  {author} {\bibinfo {author} {\bibfnamefont {R.~A.}\ \bibnamefont {Arkowitz}},\ }\href {\doibase 10.1016/S0962-8924(98)01412-3} {\bibfield  {journal} {\bibinfo  {journal} {Trends Cell Biol.}\ }\textbf {\bibinfo {volume} {9}},\ \bibinfo {pages} {20} (\bibinfo {year} {1999})}\BibitemShut {NoStop}%
\bibitem [{\citenamefont {Zigmond}(1977)}]{zigmond_ability_1977}%
  \BibitemOpen
  \bibfield  {author} {\bibinfo {author} {\bibfnamefont {S.~H.}\ \bibnamefont {Zigmond}},\ }\href {\doibase 10.1083/jcb.75.2.606} {\bibfield  {journal} {\bibinfo  {journal} {J. Cell Biol.}\ }\textbf {\bibinfo {volume} {75}},\ \bibinfo {pages} {606} (\bibinfo {year} {1977})}\BibitemShut {NoStop}%
\bibitem [{\citenamefont {Mortimer}\ \emph {et~al.}(2008)\citenamefont {Mortimer}, \citenamefont {Fothergill}, \citenamefont {Pujic}, \citenamefont {Richards},\ and\ \citenamefont {Goodhill}}]{mortimer_growth_2008}%
  \BibitemOpen
  \bibfield  {author} {\bibinfo {author} {\bibfnamefont {D.}~\bibnamefont {Mortimer}}, \bibinfo {author} {\bibfnamefont {T.}~\bibnamefont {Fothergill}}, \bibinfo {author} {\bibfnamefont {Z.}~\bibnamefont {Pujic}}, \bibinfo {author} {\bibfnamefont {L.~J.}\ \bibnamefont {Richards}}, \ and\ \bibinfo {author} {\bibfnamefont {G.~J.}\ \bibnamefont {Goodhill}},\ }\href {\doibase 10.1016/j.tins.2007.11.008} {\bibfield  {journal} {\bibinfo  {journal} {Trends Neurosci.}\ }\textbf {\bibinfo {volume} {31}},\ \bibinfo {pages} {90} (\bibinfo {year} {2008})}\BibitemShut {NoStop}%
\bibitem [{\citenamefont {Black}(2005)}]{greedy_algo_def}%
  \BibitemOpen
  \bibfield  {author} {\bibinfo {author} {\bibfnamefont {P.}~\bibnamefont {Black}},\ }\href {https://xlinux.nist.gov/dads/HTML/greedyalgo.html} {\enquote {\bibinfo {title} {greedy algorithm},}\ } (\bibinfo {year} {2nd February 2005})\BibitemShut {NoStop}%
\bibitem [{\citenamefont {Barton}(1985)}]{barton_elements_1989}%
  \BibitemOpen
  \bibfield  {author} {\bibinfo {author} {\bibfnamefont {G.}~\bibnamefont {Barton}},\ }\href {https://books.google.co.uk/books?id=NACuQgAACAAJ} {\emph {\bibinfo {title} {Elements of {Green}'s {Functions} and {Propagation}: {Potentials}, {Diffusion}, and {Waves}}}}\ (\bibinfo  {publisher} {Oxford University Press, Oxford, UK},\ \bibinfo {year} {1985})\BibitemShut {NoStop}%
\bibitem [{\citenamefont {van Haastert}\ and\ \citenamefont {Postma}(2007)}]{vanHaastert2007}%
  \BibitemOpen
  \bibfield  {author} {\bibinfo {author} {\bibfnamefont {P.~J.}\ \bibnamefont {van Haastert}}\ and\ \bibinfo {author} {\bibfnamefont {M.}~\bibnamefont {Postma}},\ }\href {\doibase 10.1529/biophysj.107.104356} {\bibfield  {journal} {\bibinfo  {journal} {Biophys. J.}\ }\textbf {\bibinfo {volume} {93}},\ \bibinfo {pages} {1787} (\bibinfo {year} {2007})}\BibitemShut {NoStop}%
\bibitem [{\citenamefont {Hein}\ \emph {et~al.}(2016)\citenamefont {Hein}, \citenamefont {Brumley}, \citenamefont {Carrara}, \citenamefont {Stocker},\ and\ \citenamefont {Levin}}]{hein_physical_2016}%
  \BibitemOpen
  \bibfield  {author} {\bibinfo {author} {\bibfnamefont {A.~M.}\ \bibnamefont {Hein}}, \bibinfo {author} {\bibfnamefont {D.~R.}\ \bibnamefont {Brumley}}, \bibinfo {author} {\bibfnamefont {F.}~\bibnamefont {Carrara}}, \bibinfo {author} {\bibfnamefont {R.}~\bibnamefont {Stocker}}, \ and\ \bibinfo {author} {\bibfnamefont {S.~A.}\ \bibnamefont {Levin}},\ }\href {\doibase 10.1098/rsif.2015.0844} {\bibfield  {journal} {\bibinfo  {journal} {J. R. Soc. Interface}\ }\textbf {\bibinfo {volume} {13}},\ \bibinfo {pages} {20150844} (\bibinfo {year} {2016})}\BibitemShut {NoStop}%
\bibitem [{\citenamefont {Greenfield}\ \emph {et~al.}(2009)\citenamefont {Greenfield}, \citenamefont {McEvoy}, \citenamefont {Shroff}, \citenamefont {Crooks}, \citenamefont {Wingreen}, \citenamefont {Betzig},\ and\ \citenamefont {Liphardt}}]{greenfield_self-organization_2009}%
  \BibitemOpen
  \bibfield  {author} {\bibinfo {author} {\bibfnamefont {D.}~\bibnamefont {Greenfield}}, \bibinfo {author} {\bibfnamefont {A.~L.}\ \bibnamefont {McEvoy}}, \bibinfo {author} {\bibfnamefont {H.}~\bibnamefont {Shroff}}, \bibinfo {author} {\bibfnamefont {G.~E.}\ \bibnamefont {Crooks}}, \bibinfo {author} {\bibfnamefont {N.~S.}\ \bibnamefont {Wingreen}}, \bibinfo {author} {\bibfnamefont {E.}~\bibnamefont {Betzig}}, \ and\ \bibinfo {author} {\bibfnamefont {J.}~\bibnamefont {Liphardt}},\ }\href@noop {} {\bibfield  {journal} {\bibinfo  {journal} {PLoS Biol.}\ }\textbf {\bibinfo {volume} {7}} (\bibinfo {year} {2009})}\BibitemShut {NoStop}%
\bibitem [{\citenamefont {Pausch}\ and\ \citenamefont {Pruessner}(2019)}]{pausch_is_2019}%
  \BibitemOpen
  \bibfield  {author} {\bibinfo {author} {\bibfnamefont {J.}~\bibnamefont {Pausch}}\ and\ \bibinfo {author} {\bibfnamefont {G.}~\bibnamefont {Pruessner}},\ }\href {\doibase 10.1088/1742-5468/ab081c} {\bibfield  {journal} {\bibinfo  {journal} {J. Stat. Mech.: Theory Exp.}\ }\textbf {\bibinfo {volume} {2019}},\ \bibinfo {pages} {053501} (\bibinfo {year} {2019})}\BibitemShut {NoStop}%
\bibitem [{\citenamefont {Athilingam}\ \emph {et~al.}(2023)\citenamefont {Athilingam}, \citenamefont {Nelanuthala}, \citenamefont {Breen}, \citenamefont {Wohland},\ and\ \citenamefont {Saunders}}]{bicoid}%
  \BibitemOpen
  \bibfield  {author} {\bibinfo {author} {\bibfnamefont {T.}~\bibnamefont {Athilingam}}, \bibinfo {author} {\bibfnamefont {A.~V.}\ \bibnamefont {Nelanuthala}}, \bibinfo {author} {\bibfnamefont {C.}~\bibnamefont {Breen}}, \bibinfo {author} {\bibfnamefont {T.}~\bibnamefont {Wohland}}, \ and\ \bibinfo {author} {\bibfnamefont {T.~E.}\ \bibnamefont {Saunders}},\ }\href {\doibase 10.1101/2022.09.28.509874} {\bibfield  {journal} {\bibinfo  {journal} {bioRxiv}\ } (\bibinfo {year} {2023}),\ 10.1101/2022.09.28.509874}\BibitemShut {NoStop}%
\bibitem [{\citenamefont {Drocco}\ \emph {et~al.}(2012)\citenamefont {Drocco}, \citenamefont {Wieschaus},\ and\ \citenamefont {Tank}}]{bicoid2}%
  \BibitemOpen
  \bibfield  {author} {\bibinfo {author} {\bibfnamefont {J.}~\bibnamefont {Drocco}}, \bibinfo {author} {\bibfnamefont {E.}~\bibnamefont {Wieschaus}}, \ and\ \bibinfo {author} {\bibfnamefont {D.}~\bibnamefont {Tank}},\ }\href {\doibase 10.1088/1478-3975/9/5/055004} {\bibfield  {journal} {\bibinfo  {journal} {Phys. Biol.}\ }\textbf {\bibinfo {volume} {9}},\ \bibinfo {pages} {055004} (\bibinfo {year} {2012})}\BibitemShut {NoStop}%
\bibitem [{\citenamefont {Lander}(2013)}]{lander_how_2013}%
  \BibitemOpen
  \bibfield  {author} {\bibinfo {author} {\bibfnamefont {A.~D.}\ \bibnamefont {Lander}},\ }\href {\doibase 10.1126/science.1224186} {\bibfield  {journal} {\bibinfo  {journal} {Science}\ }\textbf {\bibinfo {volume} {339}},\ \bibinfo {pages} {923} (\bibinfo {year} {2013})}\BibitemShut {NoStop}%
\bibitem [{\citenamefont {Parent}(1999)}]{parent_cells_1999}%
  \BibitemOpen
  \bibfield  {author} {\bibinfo {author} {\bibfnamefont {C.~A.}\ \bibnamefont {Parent}},\ }\href {\doibase 10.1126/science.284.5415.765} {\bibfield  {journal} {\bibinfo  {journal} {Science}\ }\textbf {\bibinfo {volume} {284}},\ \bibinfo {pages} {765} (\bibinfo {year} {1999})}\BibitemShut {NoStop}%
\bibitem [{\citenamefont {Berg}\ and\ \citenamefont {Brown}(1972)}]{berg_chemotaxis_1972}%
  \BibitemOpen
  \bibfield  {author} {\bibinfo {author} {\bibfnamefont {H.~C.}\ \bibnamefont {Berg}}\ and\ \bibinfo {author} {\bibfnamefont {D.~A.}\ \bibnamefont {Brown}},\ }\href {\doibase 10.1038/239500a0} {\bibfield  {journal} {\bibinfo  {journal} {Nature}\ }\textbf {\bibinfo {volume} {239}},\ \bibinfo {pages} {500} (\bibinfo {year} {1972})}\BibitemShut {NoStop}%
\bibitem [{\citenamefont {Segall}(1993)}]{segall_polarization_1993}%
  \BibitemOpen
  \bibfield  {author} {\bibinfo {author} {\bibfnamefont {J.~E.}\ \bibnamefont {Segall}},\ }\href {\doibase 10.1073/pnas.90.18.8332} {\bibfield  {journal} {\bibinfo  {journal} {Proc. Natl. Acad. Sci. Academy of Sciences}\ }\textbf {\bibinfo {volume} {90}},\ \bibinfo {pages} {8332} (\bibinfo {year} {1993})}\BibitemShut {NoStop}%
\bibitem [{\citenamefont {Bernoff}\ \emph {et~al.}(2023)\citenamefont {Bernoff}, \citenamefont {Jilkine}, \citenamefont {Navarro~Hernández},\ and\ \citenamefont {Lindsay}}]{bernoff_single-cell_2023}%
  \BibitemOpen
  \bibfield  {author} {\bibinfo {author} {\bibfnamefont {A.~J.}\ \bibnamefont {Bernoff}}, \bibinfo {author} {\bibfnamefont {A.}~\bibnamefont {Jilkine}}, \bibinfo {author} {\bibfnamefont {A.}~\bibnamefont {Navarro~Hernández}}, \ and\ \bibinfo {author} {\bibfnamefont {A.~E.}\ \bibnamefont {Lindsay}},\ }\href {\doibase 10.1016/j.bpj.2023.06.015} {\bibfield  {journal} {\bibinfo  {journal} {Biophys. J.}\ }\textbf {\bibinfo {volume} {122}},\ \bibinfo {pages} {3108} (\bibinfo {year} {2023})}\BibitemShut {NoStop}%
\bibitem [{\citenamefont {Vergassola}\ \emph {et~al.}(2007)\citenamefont {Vergassola}, \citenamefont {Villermaux},\ and\ \citenamefont {Shraiman}}]{vergassola_infotaxis_2007}%
  \BibitemOpen
  \bibfield  {author} {\bibinfo {author} {\bibfnamefont {M.}~\bibnamefont {Vergassola}}, \bibinfo {author} {\bibfnamefont {E.}~\bibnamefont {Villermaux}}, \ and\ \bibinfo {author} {\bibfnamefont {B.~I.}\ \bibnamefont {Shraiman}},\ }\href {\doibase 10.1038/nature05464} {\bibfield  {journal} {\bibinfo  {journal} {Nature}\ }\textbf {\bibinfo {volume} {445}},\ \bibinfo {pages} {406} (\bibinfo {year} {2007})}\BibitemShut {NoStop}%
\bibitem [{\citenamefont {Gerisch}(1982)}]{gerisch_chemotaxis_1982}%
  \BibitemOpen
  \bibfield  {author} {\bibinfo {author} {\bibfnamefont {G.}~\bibnamefont {Gerisch}},\ }\href {\doibase 10.1146/annurev.ph.44.030182.002535} {\bibfield  {journal} {\bibinfo  {journal} {Annu. Rev. Physiol.}\ }\textbf {\bibinfo {volume} {44}},\ \bibinfo {pages} {535} (\bibinfo {year} {1982})}\BibitemShut {NoStop}%
\bibitem [{\citenamefont {Eidi}\ \emph {et~al.}(2017)\citenamefont {Eidi}, \citenamefont {Mohammad-Rafiee}, \citenamefont {Khorrami},\ and\ \citenamefont {Gholami}}]{eidi_modelling_2017}%
  \BibitemOpen
  \bibfield  {author} {\bibinfo {author} {\bibfnamefont {Z.}~\bibnamefont {Eidi}}, \bibinfo {author} {\bibfnamefont {F.}~\bibnamefont {Mohammad-Rafiee}}, \bibinfo {author} {\bibfnamefont {M.}~\bibnamefont {Khorrami}}, \ and\ \bibinfo {author} {\bibfnamefont {A.}~\bibnamefont {Gholami}},\ }\href {\doibase 10.1039/C7SM01568B} {\bibfield  {journal} {\bibinfo  {journal} {Soft Matter}\ }\textbf {\bibinfo {volume} {13}},\ \bibinfo {pages} {8209} (\bibinfo {year} {2017})}\BibitemShut {NoStop}%
\bibitem [{\citenamefont {Alberts}\ \emph {et~al.}(2002)\citenamefont {Alberts}, \citenamefont {Johnson}, \citenamefont {Lewis}, \citenamefont {Raff}, \citenamefont {Roberts},\ and\ \citenamefont {Walter}}]{alberts_innate_2002}%
  \BibitemOpen
  \bibfield  {author} {\bibinfo {author} {\bibfnamefont {B.}~\bibnamefont {Alberts}}, \bibinfo {author} {\bibfnamefont {A.}~\bibnamefont {Johnson}}, \bibinfo {author} {\bibfnamefont {J.}~\bibnamefont {Lewis}}, \bibinfo {author} {\bibfnamefont {M.}~\bibnamefont {Raff}}, \bibinfo {author} {\bibfnamefont {K.}~\bibnamefont {Roberts}}, \ and\ \bibinfo {author} {\bibfnamefont {P.}~\bibnamefont {Walter}},\ }in\ \href {https://www.ncbi.nlm.nih.gov/books/NBK26846/} {\emph {\bibinfo {booktitle} {Molecular {Biology} of the {Cell}. 4th edition}}}\ (\bibinfo  {publisher} {Garland Science},\ \bibinfo {year} {2002})\BibitemShut {NoStop}%
\bibitem [{\citenamefont {Sender}\ \emph {et~al.}(2023)\citenamefont {Sender}, \citenamefont {Weiss}, \citenamefont {Navon}, \citenamefont {Milo}, \citenamefont {Azulay}, \citenamefont {Keren}, \citenamefont {Fuchs}, \citenamefont {Ben-Zvi}, \citenamefont {Noor},\ and\ \citenamefont {Milo}}]{sender_total_2023}%
  \BibitemOpen
  \bibfield  {author} {\bibinfo {author} {\bibfnamefont {R.}~\bibnamefont {Sender}}, \bibinfo {author} {\bibfnamefont {Y.}~\bibnamefont {Weiss}}, \bibinfo {author} {\bibfnamefont {Y.}~\bibnamefont {Navon}}, \bibinfo {author} {\bibfnamefont {I.}~\bibnamefont {Milo}}, \bibinfo {author} {\bibfnamefont {N.}~\bibnamefont {Azulay}}, \bibinfo {author} {\bibfnamefont {L.}~\bibnamefont {Keren}}, \bibinfo {author} {\bibfnamefont {S.}~\bibnamefont {Fuchs}}, \bibinfo {author} {\bibfnamefont {D.}~\bibnamefont {Ben-Zvi}}, \bibinfo {author} {\bibfnamefont {E.}~\bibnamefont {Noor}}, \ and\ \bibinfo {author} {\bibfnamefont {R.}~\bibnamefont {Milo}},\ }\href {https://www.pnas.org/doi/full/10.1073/pnas.2308511120} {\bibfield  {journal} {\bibinfo  {journal} {Proc. Natl. Acad. Sci. Academy of Sciences}\ }\textbf {\bibinfo {volume} {120}} (\bibinfo {year} {2023})}\BibitemShut {NoStop}%
\bibitem [{\citenamefont {Del~Monte}(2009)}]{del_monte_does_2009}%
  \BibitemOpen
  \bibfield  {author} {\bibinfo {author} {\bibfnamefont {U.}~\bibnamefont {Del~Monte}},\ }\href {\doibase 10.4161/cc.8.3.7608} {\bibfield  {journal} {\bibinfo  {journal} {Cell Cycle}\ }\textbf {\bibinfo {volume} {8}},\ \bibinfo {pages} {505} (\bibinfo {year} {2009})}\BibitemShut {NoStop}%
\bibitem [{\citenamefont {Krombach}\ \emph {et~al.}(1997)\citenamefont {Krombach}, \citenamefont {Münzing}, \citenamefont {Allmeling}, \citenamefont {Gerlach}, \citenamefont {Behr},\ and\ \citenamefont {Dörger}}]{krombach_cell_1997}%
  \BibitemOpen
  \bibfield  {author} {\bibinfo {author} {\bibfnamefont {F.}~\bibnamefont {Krombach}}, \bibinfo {author} {\bibfnamefont {S.}~\bibnamefont {Münzing}}, \bibinfo {author} {\bibfnamefont {A.~M.}\ \bibnamefont {Allmeling}}, \bibinfo {author} {\bibfnamefont {J.~T.}\ \bibnamefont {Gerlach}}, \bibinfo {author} {\bibfnamefont {J.}~\bibnamefont {Behr}}, \ and\ \bibinfo {author} {\bibfnamefont {M.}~\bibnamefont {Dörger}},\ }\href {\doibase 10.1289/ehp.97105s51261} {\bibfield  {journal} {\bibinfo  {journal} {Environ. Health Perspect.}\ }\textbf {\bibinfo {volume} {105}},\ \bibinfo {pages} {1261} (\bibinfo {year} {1997})}\BibitemShut {NoStop}%
\bibitem [{\citenamefont {De~Filippo}\ and\ \citenamefont {Rankin}(2020)}]{de_filippo_secretive_2020}%
  \BibitemOpen
  \bibfield  {author} {\bibinfo {author} {\bibfnamefont {K.}~\bibnamefont {De~Filippo}}\ and\ \bibinfo {author} {\bibfnamefont {S.~M.}\ \bibnamefont {Rankin}},\ }\href {https://www.frontiersin.org/articles/10.3389/fcell.2020.603230} {\bibfield  {journal} {\bibinfo  {journal} {Front. Cell Dev. Biol.}\ }\textbf {\bibinfo {volume} {8}} (\bibinfo {year} {2020})}\BibitemShut {NoStop}%
\end{thebibliography}%

\begin{widetext}
\appendix

\section{Derivation of effective velocity at finite release rate}\label{SM:veff_derivation}
In this section we derive the effective velocity of a cell in the presence of a source which releases cue particles at a finite rate $\alpha$.
We begin with Eq. (\ref{eq:delta_r_def}) from the main text:
\begin{equation}
    \overline{\Delta r_s}(r_s) = \int_0^\pi \mathrm{d}\phi \int_0^\infty \mathrm{d}\Delta t \, \Delta r_s(r_s,\phi, \Delta t) \, p(\phi, \Delta t)\;.\label{eq:delta_r_bar_def_2_sm}
\end{equation}
Expressing the joint distribustion $p(\phi,\Delta t) = p(\phi)p(\Delta t | \phi)$ as a product of Eqs. (\ref{eq:distn_3d}) and (\ref{eq:deltat_cond_prob}) in the main text,
and rewriting $\Delta r_s$ in Eq. (\ref{eq:delta_r_def}) using Eq. (\ref{eq:lambdadef2}) as
\begin{equation}
    \Delta r_s(r_s, \phi, \Delta t) = \frac{\alpha a}{\lambda(r_s, \phi, \Delta t)} - r_s\;,
\end{equation}
we arrive at
\begin{subequations}
	\begin{align}
    \overline{\Delta r}(r_s) &=  \int_0^\pi \mathrm{d} \phi p(\phi) \int_0^\infty \mathrm{d}\Delta t \: \left(\alpha a - r_s  \lambda(\Delta t) \right) \, \Exp{ - \int_0^{\Delta t} \mathrm{d} t \lambda(t)}\;, \label{eq:delta_r_bar_def_5} \\
&=  \int_0^\pi \mathrm{d} \phi p(\phi) \left( \alpha a \int_0^\infty \mathrm{d}\Delta t \: \Exp{ - \int_0^{\Delta t} \mathrm{d} t \lambda(t)} - r_s \right)\;,
    \end{align}
\end{subequations}
where we have used integration by parts for $\lambda(\Delta t)\Exp{-\int_0^{\Delta t}\plaind t' \lambda(t')}=\plaind/\plaind t \Exp{-\int_0^{\Delta t}\plaind t' \lambda(t')}$. The boundary term $\lim_{X \to \infty}\int_0^X dt \lambda(t) = +\infty$ vanishes for $\varepsilon = \alpha a/v > 1$, discussed in more detail below. Physical interpretations of this condition are discussed below and in section \ref{sec:validity_of_results} of the main text.
Since $p(\phi)$ is normalised, the expected step length becomes 
\begin{equation}\label{eq:delta_r_ave_integrals}
    \overline{\Delta r}(r_s) =  -r_s + \alpha a \int_0^\pi \mathrm{d} \phi p(\phi) \int_0^\infty \mathrm{d}\Delta t \: \Exp{ - \int_0^{\Delta t} \mathrm{d} t \lambda(t)}  \;.
\end{equation} 
The integral in the exponent can be evaluated in closed form:
\begin{subequations}
	\begin{align}
	\int_0^{\Delta t} \mathrm{d} t \lambda (t) &= \int_0^{\Delta t} dt \frac{\alpha a }{\sqrt{r_s^2 + v^2 t^2 - 2 r_s v t \cos(\phi)}}\;,\\
						   &= \frac{\alpha a}{v} \int_0^{\frac{v \Delta t}{r_s} } \frac{\mathrm{d} \beta}{\sqrt{1 + \beta^2 - 2 \beta u}} \;,\\ 
						   &= \frac{\alpha a}{v}\frac{1}{\sqrt{1 - u^2}}  \int_0^{\frac{v \Delta t}{r_s} } \frac{\mathrm{d} \beta}{\sqrt{1 + \left( \frac{\beta - u}{\sqrt{1-u^2}} \right)}} \;,\label{eq:ready_to_evaluate} 
\end{align}
\end{subequations}
where the substitutions $\beta = \frac{v t}{r_s}$, $u=\cos(\phi)$ are introduced for notational convenience.
After the substitution $q = \frac{\beta - u}{\sqrt{1 - u^2}}$ equation (\ref{eq:ready_to_evaluate}) is a standard integral and is evaluated as follows:
\begin{subequations}
   \begin{align}
	   \int_0^{\Delta t} \mathrm{d} t \lambda (t) = \frac{\alpha a}{v} \log \left(\frac{\sqrt{\big(\frac{v \Delta t}{r_s} - u\big)^2 + 1 - u^2} + \frac{ v \Delta t}{r_s} - u}{ 1 - u} \right)\;,
   \end{align}
\end{subequations}
where we have used $0\leq u^2 \leq 1$ to remove absolute value functions from the numerator and denominator of the argument of the logarithm. 
Substituting into (\ref{eq:delta_r_ave_integrals}) yields
\begin{subequations}
   \begin{align}
	   \overline{\Delta r}(r_s) &=  - r_s + \alpha a \int_{-1}^1 \mathrm{d} u \: \frac{p(u)}{(1-u)^{-\frac{\alpha a}{v} }} \int_0^\infty \mathrm{d}\Delta t \: \left(\sqrt{\bigg(\frac{v \Delta t}{r_s} - u\bigg)^2 + 1 - u^2} + \frac{v \Delta t}{r_s} - u \right)^{- \frac{\alpha a}{v} } \;,\\
				    &= - r_s +   \alpha a \int_{-1}^1 \mathrm{d} u \: \frac{p(u)}{(1-u)^{-\frac{\alpha a}{v} }} \int_{1-u}^\infty \mathrm{d}z \:\frac{r_s}{2 v}\: \left( 1 + \frac{1-u^2}{z^2} \right)z^{-\frac{\alpha a}{v} } \;,
   \end{align}
\end{subequations}
where the substitution $z = \left(\sqrt{(\frac{v \Delta t}{r_s} - u)^2 + 1 - u^2} + \frac{v \Delta t}{r_s} - u \right)$ has been used between the first and second lines. 
This integral converges for $\varepsilon = \frac{\alpha a}{v} > 1$, and yields
\begin{subequations}
   \begin{align}
	   \overline{\Delta r}(r_s) &= - r_s + \frac{\alpha a r_s}{v} \int_{-1}^1 \mathrm{d} u \: \frac{p(u)}{(1-u)^{-\frac{\alpha a}{v} }}  (1-u)^{-\frac{\alpha a}{v}}  \frac{\frac{\alpha a}{v} - u}{\left(\frac{\alpha a}{v} \right)^2 - 1} \;,\\
        &= -r_s + \frac{\alpha a r_s}{v \left(\left(\frac{\alpha a}{v} \right)^2 - 1\right)}\left(\frac{\alpha a}{v}  - \frac{a}{r_s}  \right)\;, \label{eq:delta_r_intermediate} 
   \end{align}
\end{subequations}
where the second equality uses the normalisation of $p(u)$ and the result $\int_{-1}^{1}\mathrm{d}u \: u \: p(u) = a/r_s $, which is also employed in Eqs. (\ref{eq:v_eff_3d}) in the main text.
Rearranging gives
\begin{equation}
	   \overline{\Delta r}(r_s) = \frac{\frac{\alpha a^2}{v} - r_s}{1 - \left(\frac{\alpha a}{v}\right) ^2}  \;.
\end{equation}
The expected step duration is derived similarly to the expected step length:
\begin{subequations}
   \begin{align}
    \overline{\Delta t}(r_s) &= \int_0^\pi \mathrm{d}\phi \int_0^\infty \mathrm{d}\Delta t \, p(\phi, \Delta t)\;,\label{eq:delta_r_bar_def_2_sm_2}\\
				    &= - \int_0^\pi \mathrm{d}\phi \: p(\phi) \int_0^\infty \mathrm{d}\Delta t \: \Delta t(r_s,\phi) \, \frac{\mathrm{d}}{\mathrm{d} \Delta t}  \, \Exp{ - \int_0^{\Delta t} \mathrm{d} t \lambda(t)}\;.
   \end{align}
\end{subequations}    
Integrating by parts gives
\begin{subequations}
   \begin{align}
	   \overline{\Delta t}(r_s) &= -\int_0^\pi \mathrm{d}\phi \: p(\phi) \left(\left[\Delta t \Exp {- \int_{0}^{\Delta t}dt \lambda (t)}\right]_0^\infty + \int_{0}^{\infty} \mathrm{d} \Delta t \Exp {- \int_{0}^{\Delta t}dt \lambda (t)}\right) \\
	   &= - \int_0^\pi \mathrm{d}\phi \: p(\phi) \int_{0}^{\infty} \mathrm{d} \Delta t \Exp {- \int_{0}^{\Delta t}dt \lambda (t)}\;,
   \end{align}
\end{subequations}
which is essentially Eq. (\ref{eq:delta_r_ave_integrals}), resulting in Eq. (\ref{eq:delta_r_intermediate}). 
The expected step duration is thus given by
\begin{equation}
	   \overline{\Delta t}(r_s) = \frac{\alpha r_s - v}{1 - \left(\frac{\alpha a}{v}\right) ^2}  \; \frac{a}{v^2} \;.
\end{equation}
The effective velocity of the cell, valid for $\varepsilon = \frac{\alpha a}{v}>1$, is therefore
\begin{subequations}
   \begin{align}
	   v_{\rm eff}(r_s) &= \frac{\overline{\Delta r}(r_s) }{\overline{\Delta t}(r_s) }\;, \\
			    &= v \  \frac{a^2 \alpha - r_s v}{a(\alpha r_s - v)} \;.
   \end{align}
\end{subequations}
Introducing the parameter $\varepsilon = \alpha a / v =$, the effective velocity depends on $\varepsilon$, $v$ and $a$ only:
\begin{equation}
    v_{\rm eff}(r_s) = v \  \frac{a \varepsilon - r_s}{r_s \varepsilon - a} \;.
\end{equation}

\section{Radial Transition Probability}\label{SM:radial_transition_prob}

\begin{figure}[t]
    \centering
    \includegraphics[width=0.85\textwidth]{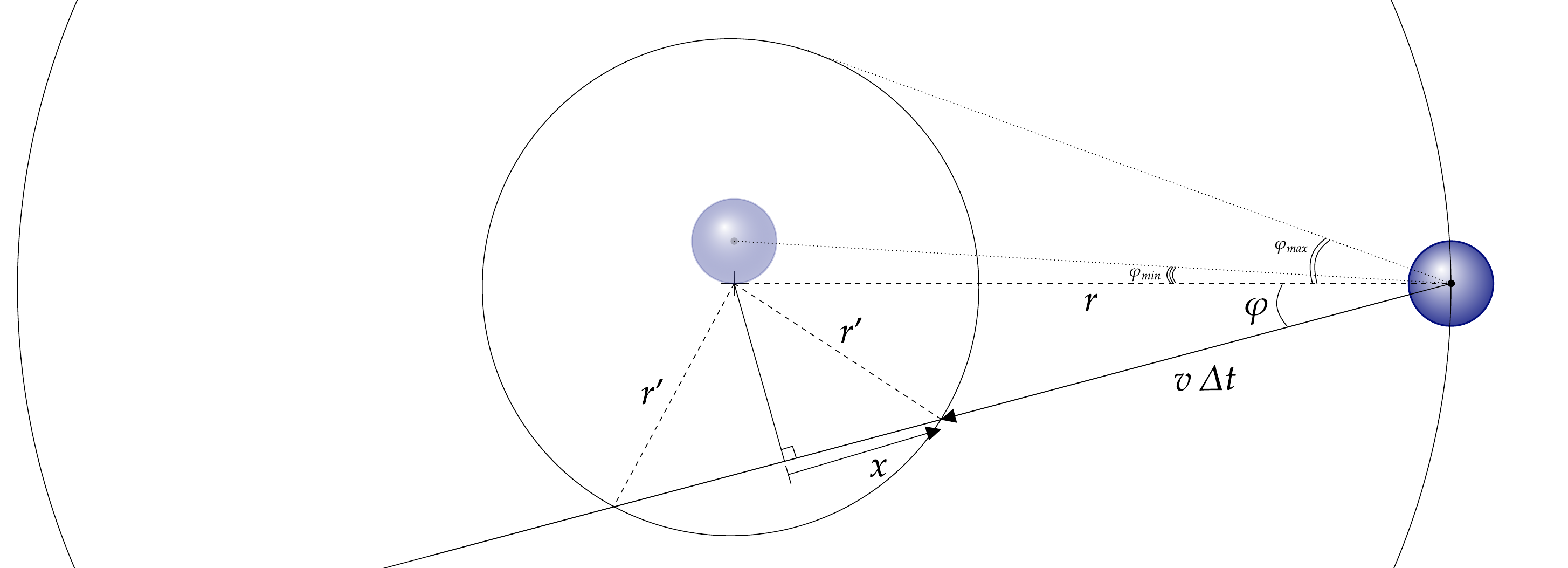}
    \caption{Schematic used in the calculation of the radial transition probability. A cell positioned at a radius $r_s = r$ from the source receives a cue particle at an angle $\phi$ with respect to the source and moves a distance $v \Delta t$, after which time it is at radius $r'$. The variable $x = v \Delta t - r \cos(\phi)$ is introduced for notational simplicity. The values of $x$ for which the transition $r \to r'$ occurs satisfy $x^2 + r^2 \sin^2(\phi)= r'^2$.  }
    \label{fig:transition}
\end{figure}

An observable of interest in the system is the success probability of a cell initialised at some distance $r_0$, \textit{i.e.} the probability that the cell reaches the source in finite time. Although calculating this in closed form is beyond the scope of this paper, we derive the transition probability $p(r \to r')$,  which denotes the probability density that a cell at a radius $r_s=r$ moves to a radius $r_s=r'$ between successive collisions with cue particles. 

The distribution of run durations, conditioned on a cue being received by a cell at $r$ at a position $u=\cos(\phi)$, is given by 
\begin{equation}
    p(\Delta t|u) = \frac{\alpha a}{r} \ \frac{1}{\sqrt{(\frac{v \Delta t}{r} - u)^2 + 1 - u^2}} \left( \frac{\sqrt{(\frac{v \Delta t}{r} - u)^2 + 1 - u^2} + \frac{v \Delta t}{r} - u}{1-u} \right)^{-\frac{\alpha a}{v}} \; .
\end{equation}
This can be cast in terms of the variable $x = v \Delta t - r u$ illustrated in figure \ref{fig:transition}:
\begin{subequations}
    \begin{align}
            p(x|u) &= p(\Delta t|u) \  \frac{\mathrm{d}\Delta t}{\mathrm{d}x}\; ,\\
            &= \frac{\alpha a}{v r} \  \frac{1}{\sqrt{(\frac{x}{r})^2 + 1 - u^2}} \left( \frac{\sqrt{(\frac{x}{r})^2 + 1 - u^2} + x}{1-u} \right)^{-\frac{\alpha a}{v}} \; .
    \end{align}
\end{subequations}
The cell makes a transition from $r \to r'$ when $x$ is a solution to $x^2 = r'^2 - r^2 \sin^2(\phi)= r'^2 - r^2(1-u^2)$. These solutions are denoted $x_+$ and $x_-$. The radial transition probability for a single run is thus given by 
\begin{equation}
    p(r \to r'|u) = p(x=x_+|u) + p(x=x_-|u)
\end{equation}
where 
\begin{subequations}
    \begin{align}
             p(x=x_\pm|u) &= \frac{\alpha a}{v r} \  \frac{1}{\sqrt{\frac{ r'^2 - r^2 (1-u^2)}{r^2} + 1 - u^2}} \left( \frac{\sqrt{\frac{ r'^2 - r^2 (1-u^2)}{r^2} + 1 - u^2} + x}{1-u} \right)^{-\frac{\alpha a}{v}} \; ,\\
             &= \frac{\alpha a}{v r'} \left( \frac{\frac{r'}{r}\pm\sqrt{\frac{r'}{r}- \sin^2(\phi)}}{1-\cos(\phi)} \right)^{-\frac{\alpha a}{v}} \; .
    \end{align}
\end{subequations}
The second equality represents a dramatic simplification of the first equality, suggesting that an analytic treatment of the transition probability may be tractable. The full transition probability valid for $r'\geq a$ is obtained by integrating over $\phi$:
\begin{subequations}
    \begin{align}
         p(r \to r') &= \int \mathrm{d}\phi \ p(r \to r'|\phi)p(\phi) \; \\
         &= 2 \int_{\phi_{\rm min}}^{\phi_{\rm max}} \mathrm{d}\phi \ \big(p(x = x_+|\phi)+ p(x = x_-|\phi)\big)p(\phi) + 2 \int_{0}^{\phi_{\rm min}} \mathrm{d} \phi \ p(x = x_-|\phi)p(\phi)\;, 
    \end{align}
\end{subequations}
where $\phi_{\rm max} = \sin ^{-1}(r'/r)$ and $\phi_{\rm min} = \sin ^{-1}(a'/r)$. The integration limits are a consequence of the fact that the process terminates when the cell surface reaches the source, so for $0<\phi <\phi_{min}$, only one value of $x$ is accessible to the cell. 
In principle, the probability of the cell reaching the source after $N$ encounters with cue particles can be written as:
\begin{equation}
    p_N(\rm success) = \int dr_1'\cdots dr_{N-1}' \; p(r \to r_1') \cdots p(r_{N-1} \to a) \;,
\end{equation}
which cannot at present be expressed in closed form.

\section{Molecular flux into cells of arbitrary shape is independent of diffusivity}\label{SM:flux_diffusivity}

Consider a cell of arbitrary shape in 3 dimensions, whose boundary is defined by a closed surface $\mathcal{S}$ enclosing a volume $\mathcal{V}$. Let $\Phi(\mathbf{r})$ be a distribution of sources and sinks of chemoattractant cue particles and assume that $\Phi(\mathbf{x})=0$ for $\mathbf{x} \in \mathcal{V}$. The density distribution of cue particles $\rho(\mathbf{r})$ is the solution of
\begin{equation}\label{eq:arbitrary_poisson_eq}
    -\nabla^2\rho(\mathbf{r}) = \frac{\alpha}{D} \; \Phi(\mathbf{r}) \fullstop
\end{equation}
This is solved by the Greens function
\begin{subequations}
    \begin{align}
        -\nabla^2 G(\mathbf{r}|\mathbf{r'}) &= \frac{\alpha}{D} \; \delta(\mathbf{r}-\mathbf{r'})\comma\\
        G(\mathbf{r}|\mathbf{r'}) &= \frac{\alpha}{D} \; \frac{1}{4\pi |\mathbf{r}-\mathbf{r'}|}\fullstop
    \end{align}
\end{subequations}
The solution to \Eref{arbitrary_poisson_eq} in the region $\vecr \notin \mathcal V $ is hence
\begin{subequations}
    \begin{align}
    \rho(\mathbf{r}) &= \int \plaind V' \; G(\mathbf{r}|\mathbf{r'}) \Phi(\mathbf{r'})\comma\\
    &= \frac{\alpha}{D} \int \plaind V' \; \frac{\Phi(\vecr')}{4\pi |\vecr - \vecr'|} \comma \label{eq:density_arbitrary_volume}
    \end{align}
\end{subequations}
and the corresponding chemoattractant gradient in the same region is 
\begin{equation}
    \mathbf{\nabla}_\vecr \rho(\vecr) = - \frac{\alpha}{D} \int \plaind V' \;\frac{\Phi(\vecr')}{4\pi} \; \frac{\vecr - \vecr'}{|\vecr - \vecr'|^3}\fullstop\label{eq:gradient_arbitrary_volume}
\end{equation}
The magnitude of the chemoattractant gradient depends on the ratio $\alpha/D$. Using Fick's Law, the chemoattractant flux into the cell at a point $\vecx \in \mathcal{S}$ is given by  
\begin{subequations}
    \begin{align}
        \vecJ (\vecx) &= -D \mathbf{\nabla}_\vecr \rho(\vecr)\bigg \rvert_{\vecr = \vecx}\\
        &= \alpha \int \plaind V' \;\frac{\Phi(\vecr')}{4\pi} \; \frac{\vecx - \vecr'}{|\vecx - \vecr'|^3}\label{eq:flux_arbitrary_volume}
    \end{align}
\end{subequations}
which is independent of the diffusivity. 
Comparing \Eref{gradient_arbitrary_volume} and \Eref{flux_arbitrary_volume}, we observe that increasing the diffusivity $D$ reduces the chemical gradient in the vicinity of the cell without changing the flux into the cell surface. This provides an explanation of the ability of cells which navigate using chemical flux to operate well in shallow gradients.




\end{widetext}

\end{document}